\documentstyle[eqsecnum,epsfig, aps,prd]{revtex}

\begin{document}
\draft
\preprint{hep-ph/} 
\title{ B-meson dileptonic decays enhanced by supersymmetry with large $\tan\beta$}
         
\author{Zhaohua Xiong $^{a,b,c}$ and Jin Min Yang $^b$} 

\address{$^a$ CCAST (World  Laboratory), P.O.Box 8730, Beijing 100080, China}
\address{$^b$ Institute of Theoretical Physics, Academia Sinica, 
           Beijing 100080, China} 
\address{$^c$ Institute of High Energy Physics, Academia Sinica,
         Beijing 100039, China}
\date{\today}
\maketitle

\begin{abstract}

We examined the rare decays $B\to X_s\ell^+\ell^-$ and 
$B_s\to\ell^+\ell^-\gamma$ in the minimal supersymmetric model with large 
$\tan\beta$. Taking into account the gluino-loop and neutralino-loop effects, we found 
that for  a large $\tan\beta$ the neutral Higgs exchanging diagrams could enhance 
$Br(B\to X_s\tau^+\tau^-)$ by a factor of 5 and $Br(B_s\to\tau^+\tau^-\gamma)$ by
a couple of orders in some part of supersymmetric parameter space allowed  by current 
experiments such as $b\to s\gamma,~B\to K^{(*)}\ell^+\ell^-$ and 
$B_s\to\ell^+\ell^-$. The  forward-backward asymmetry and the distributions of 
differential branching ratios are also found to differ significantly from 
the standard model results. Such enhanced branching ratios reach the 
level of $10^{-5}$ and thus  might be observable in the new generation of 
B experiments. 
\end{abstract} 

\pacs{12.15.Mm,  12.60.Fr,~12.60.Jv, 13.25.Hw}

\section{Introduction}
\label{sec:sec1}

Flavor-changing neutral-currents (FCNC) induced B-meson rare decays provide an ideal 
opportunity for extracting information about the fundamental parameters of the standard
model (SM), testing the SM predictions at loop level and probing possible new 
physics. After the observation of the penguin-induced decay $B\to X_s\gamma$ and the 
corresponding exclusive channels such as $B\to K^*\gamma$\cite{CLEO9395}, rare B-decays 
have begun to play an important role in the phenomenology of particle physics. The 
latest measured decay ratio for $B\to X_s\gamma$ by CLEO and BELLE\cite{CLEO01} is in
good agreement with the SM prediction, putting strong constraints on its various  
extensions and therefore stimulating the study of radiative rare B-meson decays with 
a new momentum.

Among rare B-meson decays, $B_s\to\ell^+\ell^-\gamma$ $(\ell=e,~\mu,~\tau)$  
are of special interest due to their relative cleanliness and 
sensitivity to models beyond the SM \cite{Aliev97,Iltan00,Xiong01}. Since in these 
processes a photon is emitted in additional to the lepton pair,  no helicity suppression 
exists and ``large''  branching ratio is expected.  Other interesting decay modes in this 
context are the inclusive transitions $B\to X_s\ell^+\ell^-$. Although these rare decays 
have not been observed, their detection is expected at the B-factories which 
are currently running. 

These decays have been studied in the SM\cite{Misiak93} and 
recently, to reduce the theoretical uncertainties, the next-to-next-to leading 
order(NNLO) corrections were completed \cite{Bobeta00}.
New physics effects in these decays have also been studied in some models, such as  
the minimal supersymmetric model (MSSM) \cite{Chankowski01,Grinstein97,Bertolini91,Huang99}, 
the two Higgs doublet model(2HDM) \cite{Grinstein88,Dai97,Hewett96,Logan00} 
and the technicolor model\cite{Xiong01}. 

It is noticeable that in the SM the matrix elements of $B\to X_s\ell^+\ell^-$ and 
$B_s\to\ell^+\ell^-\gamma$ are strongly suppressed by a factor $m_\ell/m_W$ and 
the contributions from exchanging neutral Higgs boson can be safely neglected.  
In the MSSM \cite{Haber85}, the situation is different, 
specially in the case of $\ell=\tau$ with large $\tan\beta$.  In this model, as 
studied in \cite{Huang99}, the contributions from  exchanging neutral Higgs
bosons are enhanced roughly by a factor $\tan^3\beta$  and may no longer be 
negligible since currently a large $\tan\beta$ is favored both 
by LEP experiments \cite{LEP_higgs} and by the supersymmetry (SUSY) explanation
\cite{Everett00} of the measurement for the muon anomalous magnetic moment 
\cite{Brown01}.

 We note that the most previous studies on the contributions from exchanging neutral
Higgs bosons \cite{Grinstein97,Bertolini91,Huang99} mainly focused on the charged-current 
loop effects; the effects of neutral-current loops (NCL), such as gluino-loop and 
neutralino-loop,  have been considered only for $B^0_{d,s}\to \mu^+\mu^-$
using mass insert approximation method in \cite{Chankowski01}.
A detail general calculation  for such NCL effects is necessary. 
The NCL can be induced via the flavor mixing of down-type 
squarks and might be important for the following reasons. 
Firstly, such flavor mixings of sfermions are almost unavoidable in supersymmetric
models. In fact, in the framework of MSSM sfermions may have arbitrary 
flavor mixings in the soft breaking terms; while in some constrained MSSM, 
such as low-energy supergravity models, the flavor mixings at weak scale could 
be naturally generated through renormalization equation even the flavor diagonality 
is assumed at the Planck scale \cite{duncan}. Secondly, the flavor mixings 
between the third and the second generation squarks are subject to no strong 
low-energy constraints like $K^0-\bar K^0$ mixing. Thirdly,  the large $\tan\beta$ 
will give rise to large mass splitting between two mass eigenstates of sbottoms, 
making the lighter sbottom ($\tilde{b}_1$) even lighter.

In this article, we will present a complete calculation of MSSM effects 
in the decays $B_s\to\ell^+\ell^-\gamma$ and  $B\to X_s\ell^+\ell^-$ 
$(\ell=e,~\mu,~\tau)$, taking into account the contributions from the neutral Higgs 
exchange with NCL. We will evaluate the effects on branching 
ratios, the forward-backward asymmetry as well as the distributions of differential 
branching ratios.  In Section \ref{sec:mixing}, we will give 
a brief description of the squark mixing in the MSSM. The detailed calculations 
for Wilson coefficients of scalar and pseudo-scalar operators are included in 
Section \ref{sec:coeff}.  The decays $B\to X_s\ell^+\ell^-$ and 
$B_s\to\ell^+\ell^-\gamma$ are calculated in Section \ref{sec:bstt}.
Experimental constraints on the parameter space of the MSSM are discussed
in Section \ref{constraints}. Some numerical results are presented in
Section \ref{sec:result}. Finally, in Section~\ref{conclusion}, we give our conclusion.

\section{Squark mixings in MSSM}
\label{sec:mixing}

The squark mass terms arise from the scalar potential which contains 
the supersymmetric F-term and D-term as well as the soft SUSY breaking 
terms. These  soft breaking terms may have arbitrary flavor mixings.
As a result, the squark mass terms in flavor basis, i.e., 
$\tilde U=(\tilde u_L, \tilde c_L,\tilde t_L, 
\tilde u_R, \tilde c_R, \tilde t_R)$ and
$\tilde D=(\tilde d_L, \tilde s_L,\tilde b_L, 
\tilde d_R, \tilde s_R, \tilde b_R)$, 
take the forms  
\begin{eqnarray}
{\tilde M}_{\tilde U}^2&=&
\left[ \begin{array}{cc}
M_{\tilde Q}^2+m_U^2+m_Z^2(\frac{1}{2}-e_U\sin^2\theta_W)\cos2\beta&m_U(A_u-\mu\cot\beta)\\
m_U(A_u-\mu\cot\beta)& M_{\tilde U}^2+m_U^2+m_Z^2e_U\sin^2\theta_W\cos2\beta  \\ 
           \end{array} \right ],\\
{\tilde M}_{\tilde D}^2&=&
\left[ \begin{array}{cc}
M_{\tilde Q}^2+m_D^2-m_Z^2(\frac{1}{2}+e_D\sin^2\theta_W)\cos2\beta&m_D(A_u-\mu\tan\beta)\\
m_D(A_d-\mu\tan\beta)& M_{\tilde D}^2+m_D^2+m_Z^2e_D\sin^2\theta_W\cos2\beta\\   
           \end{array} \right ],
\end{eqnarray}
where $e_U=2/3,~e_D=-1/3$, $m_{U(D)}$ is the $3\times 3$ diagonal mass matrix for 
up(down)-type quarks. $M_{\tilde Q}^2$, $ M_{\tilde U}^2$ and $ M_{\tilde D}^2$ are 
soft-breaking mass terms for left-handed squark doublet $\tilde Q$, right-handed up 
and down squarks, respectively. $A_u$ ($A_d$) is the coefficient of the trilinear 
term  $H_2 \tilde Q \tilde U$ ($H_1 \tilde Q \tilde D$) in soft-breaking terms and 
$\tan\beta=v_2/v_1$ is ratio of the vacuum expectation values of the two Higgs doublets. 
The hermitian matrices ${\tilde M}_{\tilde{U},\tilde{D}}^2$ can be diagonalized by 
the unitary relations, which transfer the interaction eigenstates into  the physical mass 
eigenstates. So in the general MSSM, without knowing the mechanism of SUSY breaking, 
squarks could have arbitrary flavor mixings.

However, the flavor mixings in the first two generations are subject to strong
phenomenological constraints, such as $K^0-\bar K^0$ mixing. So we only
consider the flavor mixings between the second and third generations, i.e.,
between $\tilde b$ and $\tilde s$. Further, like the analysis in \cite{Hikasa87}, 
we suppose the tree level Lagrangian is flavor diagonal and the flavor mixing 
is induced via loops. The dominant effects are from the logarithmic divergences 
caused by soft breaking terms. Such divergences must be subtracted using a soft 
counter-term at the SUSY breaking scale, such as Planck scale $M_p$. Thus a large 
logarithm factor $\ln (M_p^2/m_W^2)\approx 80$ remains after renormalization. 
In the approximation of neglecting the strange quark mass, $\tilde s_R$ 
does not mix with sbottoms. The mixing of $\tilde s_L$ with sbottoms results 
in the physical states is given by
\begin{eqnarray} \label{rotate}
\left (\begin{array}{l} 
       \tilde s_1\\ \tilde b_1\\ \tilde b_2
       \end{array} \right )
        =T^D\left(\begin{array}{l} \tilde s_L\\ \tilde b_L\\ \tilde b_R
       \end{array} \right )=\left (
             \begin{array}{ccc}
              1         &\epsilon_1^{\prime}&\epsilon_2^{\prime}\\
            \epsilon_1 &\cos\theta_b       &\sin\theta_b\\ 
            \epsilon_2 &-\sin\theta_b      &\cos\theta_b\\       
           \end{array} \right )
               \left (\begin{array}{l} \tilde s_L\\ \tilde b_L\\ \tilde b_R    
              \end{array} \right ),
\end{eqnarray}
where
\begin{eqnarray}
\epsilon_1 &=&-c_b\frac{m_t^2}{\sin^2\beta}
\left[\frac{\left(M_{\tilde Q}^2+M_{\tilde U}^2+M_{{\tilde H}_1}^2+|A_t|^2\right)
\cos\theta_b+m_bA_t^*\sin\theta_b}{m_{{\tilde b}_1}^2-
m_{{\tilde s}_1}^2}\right],\nonumber\\
\epsilon_2 &=&-c_b\frac{m_t^2}{\sin^2\beta}
\left[\frac{-\left(M_{\tilde Q}^2+M_{\tilde U}^2+M_{{\tilde H}_1}^2+|A_t|^2\right)
\sin\theta_b+m_bA_t^*\cos\theta_b}{m_{{\tilde b}_2}^2-
m_{{\tilde s}_2}^2}\right],\nonumber\\
\epsilon_1^{\prime}&=&-\epsilon_1 \cos\theta_b+\epsilon_2 \sin\theta_b,\nonumber\\
\epsilon_2^{\prime}&=&-\epsilon_1 \sin\theta_b-\epsilon_2 \cos\theta_b.
\end{eqnarray}
Here $\theta_b$ is the left-right sbottom mixing angle,
$K_{ij}$ are the CKM-matrix elements, and $c_b=\frac{\alpha_{em}}{4\pi}\frac{K^*_{tb}
K_{ts}}{2m_W\sin^2\theta_W}\ln\frac{M_p^2}{m_W^2}$.
 
Note the mixing matrix $T^U$ between the left-handed scharm and stops takes
the similar form as $T^D$, and under the assumption that the flavor 
mixing between the second and third generation squarks is at least one order 
lower than the third left- and right-hand squarks mixing, the rotation matrix 
$T^{U,D}$ are approximately unitarity.

\section{Calculation of scalar and pseudo-scalar Wilson coefficients}
\label{sec:coeff}
In the MSSM, the short distance contribution to 
$b\to s\ell^+\ell^-$ decay can be computed in the framework of the QCD corrected 
effective weak Hamiltonian, obtained by integrating out heavy particles, i.e., 
top quark, $W^\pm,\ Z$ bosons in the SM and the sparticles,
\begin{equation}
{\cal H}_{eff}=-\frac{4G_F}{\sqrt{2}}K_{tb}K_{ts}^*
\sum_{i=1}^{10}\left [C_i(\mu_r)
{\cal O}_i(\mu_r)+C_{Q_i}(\mu_r){\cal Q}_i(\mu_r)\right ]\ ,
\label{hamilton}
\end{equation}
where ${\cal O}_i,~{\cal Q}_i$ are operators given in Eqs.~(\ref{operator1})
\cite{Grinstein88}, (\ref{operator2})\cite{Dai97} and $C_i, C_{Q_i}$ are Wilson 
coefficients renormalized at the scale $\mu_r$
\footnote{The most general Hamiltonian in 
low-energy supersymmetry also contains the operators ${\cal O}_i^\prime,
~{\cal Q}_i^\prime$ which are flipped chirality partners of ${\cal O}_i,
~{\cal Q}_i$. However, they give negligible contributions and thus not considered 
in the final discussion of physical quantities\cite{Lunghi99}.}.

\begin{figure}[htb]
\begin{center}
\epsfig{file=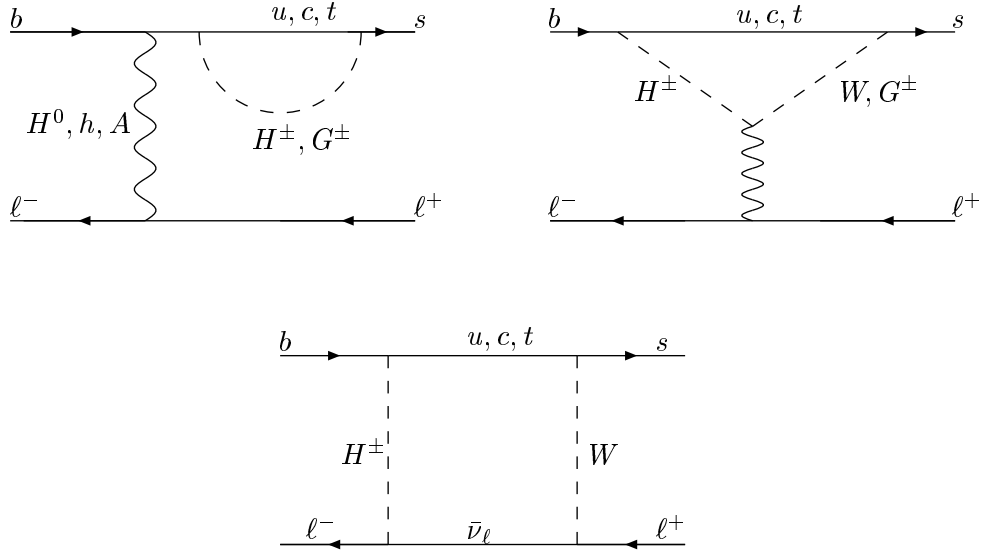,width=14cm}
\caption{The Feynman diagrams involving charged Higgs which give dominate 
contribution to $C_{Q_{1,2}}$}
\label{feyman1}
\end{center}
\end{figure}

In the MSSM the additional contributions to operators in Eq.~(\ref{hamilton})
can be  characterized by the values of the coefficients $C_{i}$ 
and $C_{Q_i}$ at the perturbative scale $m_W$. For the processes we will study, 
it is only relevant with the effective Wilson coefficients $C_{7,9,10}$ 
which have been computed in Ref.\cite{Bertolini91} and $C_{Q_{1,2}}$
of additional scalar and pseudo-scalar operators.
In this section we will focus our attention on the calculation of the Wilson 
coefficients $C_{Q_{1,2}}(m_W)$ with the assumption that except 
for the third generation squarks all sfermions are degenerate and have a mass of 
$\sim 1$ TeV. Besides the contribution from  box diagrams,  
the neutral Higgs-bosons exchange diagrams include totally five classes of loops: 
(1) W boson and up-type quarks,
(2) charged Higgs and up-type quarks, (3) charginos and up-type squarks, 
(4) neutralinos and down-type squarks, and (5) gluinos and down-type squarks.
As pointed out in Section~\ref{sec:sec1}, the last two classes of loops have 
not been calculated in the literatures. Now we take 
all of them into account and use the Feynman rules presented in 
\cite{Haber85}. Since we are only interested in large $\tan\beta$ case,  
for simplicity we ignore less important terms and keep only the leading part 
contributions given in the following: 
\begin{itemize}
\item Charged Higgs (Fig.~\ref{feyman1}) 
\end{itemize}

\begin{equation}
C_{Q_1}^{H^\pm}(m_W)=-C_{Q_2}^{H^\pm}(m_W)
=\frac{m_\ell m_b}{m_W^2}\frac{\tan^2\beta}{4}
P_1\left(x_{H^\pm t}\right)
\end{equation}
where $x_{ij}=m_i^2/m_j^2$ and the one-loop integral functions $P_i$ are
given in the appendix.
\begin{figure}[htb]
\begin{center}
\epsfig{file=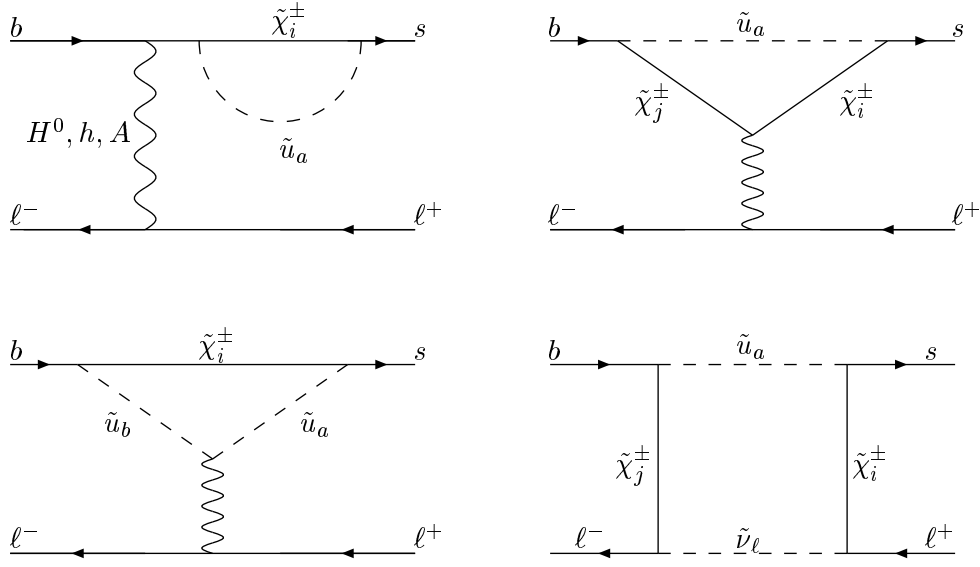,width=14cm}
\caption{The Feynman diagrams involving chargino which give 
 dominate contribution to $C_{Q_{1,2}}$}
\label{feyman2}
\end{center}
\end{figure}
\begin{itemize}
\item Chargino (Fig.~\ref{feyman2}) 
\end{itemize}

\begin{eqnarray}
C_{Q_1}^{\tilde{\chi}^\pm}(m_W)&=&-
\frac{m_\ell m_b}{4}\tan^2\beta
\sum\limits_{i,i^\prime=1}^2\sum\limits_{k,k^\prime=1}^3
\Gamma_1(i,k,k^\prime)U_{i^\prime2}\left\{
\delta_{ii^\prime}\delta_{kk^\prime}\frac{\sqrt{2}}{\cos\beta}
\frac{m_{\tilde{\chi}^\pm_i}}{m_W}
r_+P_1\left(x_{\tilde{\chi}_i^\pm\tilde{t}_{k-1}}\right)
\right.\nonumber\\
&&\left.-\delta_{kk^\prime}\Gamma_2^*(i^\prime,i)
r_+P_2\left(x_{\tilde{t}_{k-1}\tilde{\chi}_i^\pm},
x_{\tilde{\chi}^\pm_{i^\prime}\tilde{\chi}_i^\pm}\right)
-\delta_{kk^\prime}\frac{m_{\tilde{\chi}^\pm_{i^\prime}}}
{m_{\tilde{\chi}_i^\pm}}\Gamma_2(i,i^\prime)
P_3\left(x_{\tilde{t}_{k-1}\tilde{\chi}_i^\pm},
x_{\tilde{\chi}_{i^\prime}^\pm\tilde{\chi}_i^\pm}\right)
\right.\nonumber\\
&&\left.+\delta_{ii^\prime}\sqrt{2}\frac{m_W}{m_{\tilde{\chi}_i^\pm}}
\left(r_0\frac{\frac{1}{2}-\frac{2}{3}\sin^2\theta_W}{\cos^2\theta_W}
-\Gamma_3(k,k^\prime)\right)
P_3\left(x_{\tilde{t}_{k-1}\tilde{\chi}_i^\pm},
x_{\tilde{t}_{k^\prime-1}\tilde{\chi}_i^\pm}\right)
\right.\nonumber\\
&&\left.+\delta_{kk^\prime}
\frac{m_{\tilde{\chi}_{i^\prime}}}{m^3_{\tilde{\chi}_i^\pm}}
V_{i1}U_{i^\prime2}
P_4\left(x_{\tilde{m}\tilde{\chi}_i^\pm},
x_{\tilde{\chi}_{i^\prime}^\pm\tilde{\chi}_i^\pm}
,x_{\tilde{t}_{k-1}\tilde{\chi}_i^\pm}\right)
\right\},\\
C_{Q_2}^{\tilde{\chi}^\pm}(m_W)&=&
\frac{m_\ell m_b}{4m_A^2}\tan^2\beta
\sum\limits_{i,i^\prime=1}^2\sum\limits_{k,k^\prime=1}^3
\Gamma_1(i,k,k^\prime)U_{i^\prime2}\left\{
\delta_{ii^\prime}\delta_{kk^\prime}\frac{\sqrt{2}}{\cos\beta}
\frac{m_{\tilde{\chi}^\pm_i}}{m_W}
P_1\left(x_{\tilde{\chi}_i^\pm\tilde{t}_{k-1}}\right)
\right.\nonumber\\
&&\left.-2\delta_{kk^\prime}V^*_{i^\prime1}U_{i2}
P_2\left(x_{\tilde{t}_{k-1}\tilde{\chi}_i^\pm},
x_{\tilde{\chi}^\pm_{i^\prime}\tilde{\chi}_i^\pm}\right)
+2\delta_{kk^\prime}V_{i1}U_{i^\prime2}
\frac{m_{\tilde{\chi}^\pm_{i^\prime}}}{m_{\tilde{\chi}_i^\pm}}
P_3\left(x_{\tilde{t}_{k-1}\tilde{\chi}_i^\pm},
x_{\tilde{\chi}^\pm_{i^\prime}\tilde{\chi}_i^\pm}\right)
\right.\nonumber\\
&&\left.+\delta_{ii^\prime}\sqrt{2}\frac{m_t}{m_W}
\frac{\mu}{m_{\tilde{\chi}_i^\pm}}
(T^U_{k2}T^U_{k^\prime3}-T^U_{k3}T^U_{k^\prime2})
P_3\left(x_{\tilde{t}_{k-1}\tilde{\chi}_i^\pm},
x_{\tilde{t}_{k^\prime-1}\tilde{\chi}_i^\pm}\right)
\right.\nonumber\\
&&\left.-\delta_{kk^\prime}x_{A\tilde{\chi}_i^\pm}
\frac{m_{\tilde{\chi}_{i^\prime}}}{m_{\tilde{\chi}_i^\pm}}
V_{i1}U_{i^\prime2}P_4\left(x_{\tilde{m}\tilde{\chi}_i^\pm},
x_{\tilde{\chi}_{i^\prime}^\pm\tilde{\chi}_i^\pm}
,x_{\tilde{t}_{k-1}\tilde{\chi}_i^\pm}\right)
\right\},
\end{eqnarray}
where  $\alpha$ is the mixing angle of neutral components of the two Higgs 
doublets, $r_+=\cos^2\alpha m_H^{-2}+\sin^2\alpha m_h^{-2}$, 
$r_0=\sin 2\alpha(m_H^{-2}-m_h^{-2})$
$\lambda_t=m_t/{\sqrt{2}m_W\sin\beta}$ and 
$\lambda_b=m_b/{\sqrt{2}m_W\cos\beta}$ are the Yukawa couplings of top and 
bottom quarks, respectively. 
$U$, $V$ and $N$ are the matrices which diagonalise the chargino and
neutralino mass matrices. $\tilde{m}=m_{\tilde{t}_0}
=m_{\tilde{b}_0}$ is defined as the common  
mass of the first two generation squarks and all sleptons which are assumed 
to be degenerate, $N_j^\prime=\frac{1}{3}\tan\theta_WN_{j1}-N_{j2}$, and 
\begin{eqnarray}
\Gamma_1(i,k,k^\prime)&=&\left\{
\begin{array}{l} 
-V_{i1}(\frac{K_{cs}^*}{K_{ts}^*}T^U_{12}-1)~~~~~~~~~for~~k,k^\prime=1\\
\left[-V_{i1}\left(T^U_{k1}\frac{K^*_{cs}}{K_{ts}^*}+T^U_{k2}\right)
+\lambda_tV_{i2}T^U_{k3}\right]T^U_{k^\prime 2},\nonumber\\
\end{array}\right.\nonumber\\
\Gamma_2(i,i^\prime)&=&2r_+V_{i1}U_{i^\prime2}
+r_0V_{i2}U_{i^\prime1},\nonumber\\
\Gamma_3(k,k^\prime)&=&\frac{m_t(r_0A_t+2r_0m_t+2r_+\mu)}{2m_W^2}
(T^U_{k2}T^U_{k^\prime3}+T^U_{k3}T^U_{k^\prime2})
-r_0\frac{2}{3}\tan^2\theta_WT^U_{k3}T^U_{k^\prime3},\nonumber\\
\Gamma_4(j,k)&=&\frac{2}{3}\tan\theta_WN_{j1}T_{k3}^D+\sqrt{2}\lambda_b
N_{j3}^*T_{k2}^D,\nonumber\\
\Gamma_5(j,j^\prime,j^{\prime\prime})
&=&\frac{1}{2}N_{j^\prime3}\left[N_{jj^{\prime\prime}}
(N_{j^\prime2}-N_{j^\prime1}\tan\theta_W)+N_{j^\prime j^{\prime\prime}}
(N_{j2}-N_{j1}\tan\theta_W)\right],\nonumber\\
\Gamma_6(k,k^\prime,j)&=&r_0N_{j3}*T^D_{k^\prime2}\left(
\frac{\frac{1}{2}-\frac{1}{3}\sin^2\theta_W}{\cos^2\theta_W}
-\frac{1}{3}\tan^2\theta_WT^D_{k3}T^D_{k^\prime3}\right)\nonumber\\
&&+\frac{r_0\mu+2r_+A_b}{3m_W}\tan\theta_WT^D_{k^\prime3}
(T^D_{k2}T^D_{k^\prime3}+T^D_{k3}T^D_{k^\prime2}),\nonumber\\
\Gamma_7(k,j,j^\prime)&=&\left[N_{j3}(N_{j^\prime2}+\tan\theta_WN_{j^\prime1})-
2\tan\theta_WN_{j1}N_{j^\prime3}\right]N_{j^\prime3}T^D_{k2},\nonumber\\
\Gamma_8(k,j,j^\prime)&=&2\tan\theta_W(N_{j3}^*
(N_{j^\prime1}^*-N_{j1}N_{j^\prime3})N_{j3}T^D_{k3}.
\end{eqnarray}

\begin{figure}[htb]
\begin{center}
\epsfig{file=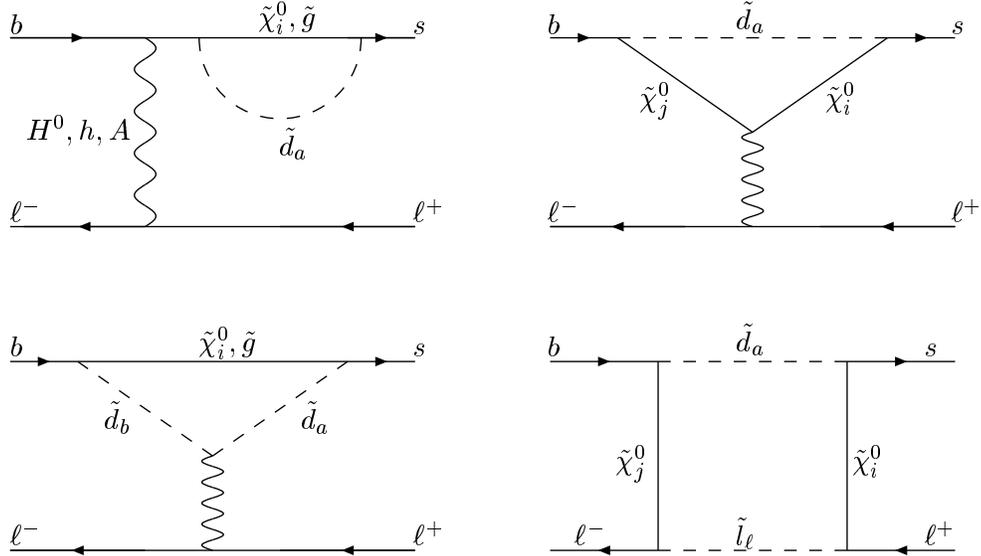,width=14cm}
\caption{The Feynman diagrams involving neutralino which give dominate contribution 
to $C_{Q_{1,2}}$}
\label{feyman3}
\end{center}
\end{figure}
\begin{itemize}
\item Neutralino (Fig.~\ref{feyman3}) 
\end{itemize}
The formula for the contribution from the neutralino are given by 
\begin{eqnarray}
C_{Q_1}^{\tilde{\chi}^0}(m_W)&=&
\frac{m_\ell m_b}{4}\frac{1}{K_{ts}^*K_{tb}}
\tan^2\beta\sum\limits_{j,j^\prime=1}^4\sum\limits_{k,k^\prime=1}^3
N_j^\prime T_{k1}^D\left\{
\delta_{jj^\prime}\delta_{kk^\prime}
\frac{m_{\tilde{\chi}_j^0}}{m_b}\Gamma_4(j,k)r_+
P_1\left(x_{\tilde{\chi}_j^0\tilde{b}_{k-1}}\right)
\right.\nonumber\\
&&\left.+\delta_{kk^\prime}\left(2r_+\Gamma^*_5(j^\prime,j,3)-
r_0\Gamma^*_5(j^\prime,j,4)\right)T^D_{k2}
P_2\left(x_{\tilde\chi^0_j\tilde{b}_{k-1}},
x_{\tilde\chi^0_{j^\prime}\tilde{b}_{k-1}}\right)
\right.\nonumber\\
&&\left.+\delta_{kk^\prime}\frac{m_{\tilde\chi^0_{j^\prime}}}{m_{\tilde\chi^0_j}}
\left(2r_+\Gamma_5(j,j^\prime,3)-r_0\Gamma_5(j,j^\prime,4)\right)
T^D_{k2}P_3\left(x_{\tilde{b}_{k-1}\tilde\chi^0_j}
x_{\tilde\chi^0_{j^\prime}\tilde\chi^0_j}\right)
\right.\nonumber\\
&&\left.-\delta_{jj^\prime}\frac{m_W}{m_{\tilde{\chi}^0_j}}
\Gamma_6(k,k^\prime,j)P_3\left(x_{\tilde{b}_{k-1}\tilde{\chi}_j^0},
x_{\tilde{b}_{k^\prime-1}\tilde{\chi}_j^0}\right)\right.\nonumber\\
&&\left.-\delta_{kk^\prime}\frac{1}{2}\frac{m_{\tilde{\chi}^0_{j^\prime}}}
{m^3_{\tilde{\chi}^0_j}}
\left(\Gamma_7(k,j,j^\prime)+\Gamma_8(k,j,j^\prime)\right)
P_4\left(x_{\tilde{m}\tilde{\chi}^0_j}
,x_{\tilde{\chi}^0_{j^\prime}\tilde{\chi}^0_j}
x_{\tilde{b}_{k-1}\tilde{\chi}^0_j}\right)\right\},\\
C_{Q_2}^{\tilde{\chi}^0}(m_W)&=&
-\frac{m_\ell m_b}{4m_A^2}\frac{1}{K_{ts}^*K_{tb}}
\tan^2\beta\sum\limits_{j,j^\prime=1}^4\sum\limits_{k,k^\prime=1}^3
N_j^\prime T_{k1}^D\left\{
\delta_{jj^\prime}\delta_{kk^\prime}\frac{m_{\tilde{\chi}_j^0}}{m_b}\Gamma_4(j,k)
P_1\left(x_{\tilde{\chi}_j^0\tilde{b}_{k-1}}\right)
\right.\nonumber\\
&&\left.-2\delta_{kk^\prime}\Gamma^*_5(j^\prime,j,3)T^D_{k2}
P_2\left(x_{\tilde\chi^0_j\tilde{b}_{k-1}},
x_{\tilde\chi^0_{j^\prime}\tilde{b}_{k-1}}\right)
+2\delta_{kk^\prime}\frac{m_{\tilde\chi^0_{j^\prime}}}{m_{\tilde\chi^0_j}}
\Gamma_5(j,j^\prime,3)T^D_{k2}
P_3\left(x_{\tilde{b}_{k-1}\tilde\chi^0_j}
x_{\tilde\chi^0_{j^\prime}\tilde\chi^0_j}\right)
\right.\nonumber\\
&&\left.-\delta_{jj^\prime}
\left(\frac{\mu}{m_{\tilde{\chi}^0_j}}T^D_{k^\prime2}-
\frac{2A_b}{3m_{\tilde{\chi}^0_j}}\tan\theta_WT^D_{k^\prime3}\right)
N_{j1}P_3\left(x_{\tilde{b}_{k-1}\tilde{\chi}_j^0},
x_{\tilde{b}_{k^\prime-1}\tilde{\chi}_j^0}\right)\right.\nonumber\\
&&\left.+\delta_{kk^\prime}\frac{1}{2}x_{A\tilde{\chi}^0_j}
\frac{m_{\tilde{\chi}^0_{j^\prime}}}{m_{\tilde{\chi}^0_j}}
\left(\Gamma_7(k,j,j^\prime)-\Gamma_8(k,j^\prime,j)\right)
P_4\left(x_{\tilde{m}\tilde{\chi}^0_j}
,x_{\tilde{\chi}^0_{j^\prime}\tilde{\chi}^0_j}
x_{\tilde{b}_{k-1}\tilde{\chi}^0_j}\right)\right\}.
\end{eqnarray}  

\begin{itemize}
\item Gluino (Fig.~\ref{feyman3}) 
\end{itemize}
\begin{eqnarray}
C_{Q_1}^{\tilde{g}}(m_W)&=&-
\frac{m_\ell m_b}{4}\frac{16g_s^2}{3g^2K_{ts}^*K_{tb}}
\tan^2\beta\sum\limits_{k,k^\prime=1}^3T^D_{k1}T^D_{k^\prime3}
\left[\delta_{kk^\prime}r_+\frac{m_{\tilde{g}}}{m_b}
P_1(x_{\tilde{g}\tilde{b}_{k-1}})\right.\nonumber\\
&&\left.+\frac{r_0\mu+2r_+A_b}{2m_{\tilde{g}}}
(T^D_{k2}T^D_{k^\prime3}+T^D_{k3}T^D_{k^\prime2})
P_3(x_{\tilde{b}_{k-1}\tilde{g}},x_{\tilde{b}_{k^\prime-1}\tilde{g}})\right],
\nonumber\\
C_{Q_2}^{\tilde{g}}(m_W)&=&
\frac{m_\ell m_b}{4m_A^2}\frac{16g_s^2}{3g^2K_{ts}^*K_{tb}}
\tan^2\beta\sum\limits_{k,k^\prime=1}^3T^D_{k1}T^D_{k^\prime3}
\left[\delta_{kk^\prime}\frac{m_{\tilde{g}}}{m_b}
P_1(x_{\tilde{g}\tilde{b}_{k-1}})\right.\nonumber\\
&&\left.+\frac{A_b}{m_{\tilde{g}}}(T^D_{k2}T^D_{k^\prime3}-T^D_{k3}T^D_{k^\prime2})
P_3(x_{\tilde{b}_{k-1}\tilde{g}},x_{\tilde{b}_{k^\prime-1}\tilde{g}})\right].
\end{eqnarray}

It is noticeable that the supersymmetric contributions to $C_{Q_{1,2}}$ have an 
overall enhancement factor $\tan^2\beta$. Moreover, the gluino loop contributions 
have an additional enhancement factor $\frac{16g_s^2}{3g^2K_{tb}K_{ts}^*}\frac{m_{\tilde g}}{m_b}$.
So sizable contributions from neutral Higgs penguin diagrams are expected for a 
sufficiently large $\tan\beta$.

\section{B rare dileptonic decays in the MSSM}
\label{sec:bstt}

\subsection{Inclusive decay $B\to X_s\ell^+\ell^-$}

Neglecting the strange quark mass, and with $p$ standing for the momentum transfer, 
the effective Hamiltonian (\ref{hamilton}) leads to the following matrix element for 
the inclusive $b\to s\ell^+\ell^-$ decay,
\begin{eqnarray}
{\cal M}&=&\frac{\alpha_{em}G_F}{2\sqrt{2}\pi}K_{tb}K_{ts}^*
\left\{-2C_7^{eff}\frac{m_b}{p^2}\bar{s}i\sigma_{\mu\nu}p_\nu(1+\gamma_5)b
\right.\nonumber\\
&&\left.+C_9^{eff}\bar{s}\gamma_\mu(1-\gamma_5) b\bar{\ell}\gamma_\mu\ell+
C_{10}\bar{s}\gamma_\mu(1-\gamma_5) b\bar{\ell}\gamma_\mu\gamma_5\ell
\right.\nonumber\\
&&\left.+C_{Q_1}\bar{s}(1+\gamma_5)b\bar{\ell}\ell
+C_{Q_2}\bar{s}(1+\gamma_5)b\bar{\ell}\gamma_5\ell\right\}.
\label{matrix}
\end{eqnarray}

The Wilson coefficients can be evaluated from $m_W$ down to the lower scale of about 
$m_b$ by using the renormalization group equation. When evolving down to b quark 
scale, the operators ${\cal O}_j~(j=1-6),~Q_3$ can mix with ${\cal O}_i,\  (i=7,9)$; 
however, they can be included in an ``effective'' ${\cal O}_{7,9}$ because of their 
same structures contributing to the  $b\to s\ell^+\ell^-$ matrix element.  
As for long-distance contribution from the intermediate 
$J/\Psi$ family, we follow Ref.\cite{Ali91} and 
include the effect in ``effective'' $C_9^{eff}$. Expanding 
$C_i$ in powers of $\alpha_s$, i.e., $C=C^0+\frac{\alpha_s}{4\pi}C^1$, 
one obtains the leading order effective 
Wilson coefficients\cite{Grinstein88,Dai97,Hewett96}   
\begin{eqnarray} 
C_7^{0,eff}(m_b)&=&\eta^{16/23}\left[C_7^{0,eff}(m_W)+\frac{8}{3}(\eta^{-2/23}-1)
C_8^{0,eff}(m_W)-0.012C_{Q_3}(m_W)\right]
+\sum\limits_{i=1}^8h_i\eta^{a_i},\\
\label{c70}
C_9^{0,eff}(m_b)&=&C_9^{0,eff}(m_W)
+\frac{2}{9}\left[3C_3+C_4+3C_5+C_6\right]\nonumber\\
&&-\frac{1}{2}g\left(1,s\right)
\left[4C_3+4C_4+3C_5+C_6\right]-\frac{1}{2}g\left(0,s\right)
\left[C_3+3C_4\right]\nonumber\\
&&+\left\{g\left(\frac{m_c^2}{m_b^2},s\right)
-\frac{3\pi}{\alpha_{em}^2}\kappa\sum_{V_i=\Psi^{'},\Psi^{''},\cdots}
\frac{m_{V_i}\Gamma(V_i\to \ell^+\ell^-)}{m_{V_i}^2-p^2-im_{V_i}\Gamma_{V_i}}\right\}
\nonumber \\
\label{c90}
&&\times\left[3C_1+C_2+3C_3+C_4+3C_5+C_6\right],\\
C_{10}^0(m_b)&=&C_{10}(m_W),\\
C_{Q_i}(m_b)&=&\eta^{-12/23}C_{Q_i}(m_W).
\end{eqnarray}
Here $s=p^2/m_b^2$ is the scaled dilepton invariant mass square, 
, $\eta=\alpha_s(m_W)/\alpha_s(m_b)$, vector $h_i, a_i, f_i$ 
and $g_i$ are given in \cite{Chetyrkin97} and 
\begin{eqnarray}
C_{Q_3}(m_W)&=&\frac{m_b}{m_\tau}(C_{Q_1}(m_W)+C_{Q_2}(m_W)),\\
C_{1-6}&=&\left(-0.4561,1.0208,-0.0041,-0.0603,0.0028,0.0037\right).
\end{eqnarray}
At NLO level, the Wilson coefficient $C_4^{1,eff}$ can be found 
in \cite{Chetyrkin97} and 
\begin{eqnarray} 
C_7^{1,eff}(m_b)&=&\frac{37208}{4761}\eta^{16/23}(\eta-1)C_7^{0,eff}(m_W) +
\eta^{39/23}C_7^{1,eff}(m_W)+\frac{8}{3}\eta^{37/23}(1-\eta^{2/23}))C_8^{1,eff}(m_W)
\nonumber\\
&&+\frac{4}{14283}\eta^{14/23}
\left[64217\eta+74416\eta^{2/23}-\frac{1791104}{25}-\frac{1674721}{25}\eta^{25/23}\right]
C_8^{0,eff}\nonumber\\
&&+\sum\limits_{i=1}^8\left[e_i\eta C_4^{1,eff}(m_W)+f_i+g_i\eta\right]\eta^{a_i},\\
C_9^{1,eff}(m_b)&=&\frac{1}{12}C_9^{0,eff}(m_W)\left[-4Li_2(s)-2\ln(s)\ln(1-s)
-\frac{2}{3}\pi^2-\frac{5+4s}{1+2s}\ln(1-s)\right.\nonumber\\
&&\left.-\frac{2s(1+s)(1-2s)}{(1-s)^2(1+2s)}\ln(s)+\frac{5+9s-6s^2}{2(1-s)(1+2s)}\right].
\end{eqnarray}
Function $g(m_c^2/m_b^2,s)$ in Eq.\ (\ref{c90}) arises from 
the one-loop matrix elements of the four-quark operators, and
\begin{equation}
g(x,y)=-\frac{4}{9}\ln\ x +\frac{8}{27}+\frac{16x}{9y}-
\frac{4}{9}(1+\frac{2x}{y}){\vert 1-\frac{4x}{y}\vert}^{1/2}
\left\{
\begin{array}{cc}
\ln\ Z(x,y)-i\pi & for~~4x/y< 1\nonumber\\
2\arctan \frac{1}{\sqrt{4x/y-1}},&for~~4x/y> 1
\end{array}
\right.                                      
\end{equation}
where
\begin{equation}
Z(x,y)=\frac{1+\sqrt{1-\frac{4x}{y}}}{1-\sqrt{1-\frac{4x}{y}}} .
\label{zxy}
\end{equation}
To  estimate the long-distance contribution in  the second term in brace of 
Eq.\ (\ref{c90}), we take the phenomenological parameter $\kappa$ as 2.3\cite{Ali91} in 
numerical calculations.

The formula of invariant dilepton mass distribution has been derived
in \cite{Dai97}, which is given by
\begin{eqnarray}\label{dgbtt}
\frac{d\Gamma(B\to X_s\ell^+\ell^-)}{ds}
=\frac{G_F^2m_b^5}{768\pi^5}\alpha_{em}^2\vert K_{tb}K_{ts}^*\vert^2
(1-s)^2(1-\frac{4r}{s})^{1/2}D(s)
\label{bll}
\end{eqnarray}
with 
\begin{eqnarray}
D(s)&=&4|C_7^{eff}|^2(1+\frac{2r}{s})(1+\frac{2}{s})+
|C_9^{eff}|^2(1+\frac{2r}{s})(1+2s)+|C_{10}|^2
(1-8r+2s+\frac{2r}{s})\nonumber\\
&&+12Re(C_7^{eff}C_9^{eff*})(1+\frac{2r}{s})
+\frac{3}{2}|C_{Q_1}|^2(s-4r)+\frac{3}{2}|C_{Q_2}|^2s
+6Re(C_{10}C_{Q_2}^*)r^{1/2}
\label{bllp}
\end{eqnarray}
where $r=m_\ell^2/m_b^2$. To get rid of large uncertainties due to $m_b^5$
and CKM elements in Eq.~(\ref{bll}), we normalized the decay rate to the 
semileptonic decay rate 
\begin{equation}
\Gamma(B\to X_c\ell\nu)=\frac{G_F^2m_b^5}{192\pi}\vert K_{cb}\vert^2
f(\frac{m_c}{m_b})k(\frac{m_c}{m_b}).
\end{equation}
Here $f(x)=1-8x^2+8x^6-x^8-24x^4\ln x$ is the phase-space factor and 
$k(x)$ is a sizable next-to-leading QCD correction to the semileptonic decay
\cite{Cabibbo78}.

The angular information and the forward-backward (FB) asymmetry are also 
sensitive to the details of the new physics. Defining the forward-backward  
asymmetry as
\begin{equation}
A_{FB}(s)=\frac{\int_0^1 d\cos\theta(d^2\Gamma/dsd\cos\theta)
-\int_{-1}^0 d\cos\theta(d^2\Gamma/dsd\cos\theta)}
{\int_0^1 d\cos\theta (d^2\Gamma/dsd\cos\theta)+\int_{-1}^0 d\cos\theta
(d^2\Gamma/dsd\cos\theta)} \ , 
\label{af}
\end{equation}    
where $\theta$ is the angle between the momentum of B-meson and 
$\ell^+$ in the center of mass frame of the dilepton, we obtain
\begin{equation}
A_{FB}=\frac{6(1-4r/s)^{1/2}}{D(s)}Re\left[2C_7^{eff}C_{10}^*+C_9^{eff}C_{10}^*s
+ 2C_7^{eff}C_{Q_2}^*r^{1/2}+C_9^{eff}C_{Q_1}^*r^{1/2}\right].
\end{equation}

\subsection{Exclusive decay $B_s\to \ell^+\ell^-\gamma$}

Now let us turn to the rare radiative decay  
$B_s\to \ell^+\ell^-\gamma$. The  exclusive decay can be obtained 
from the inclusive decay $b\to s\ell^+\ell^-\gamma$, and further, 
from $b\to s\ell^+\ell^-$. To achieve this, it is necessary to attach
photon to any charged internal and external lines in the Feynman diagrams
of $b\to s\ell^+\ell^-$. As pointed out in Ref.\ \cite{Aliev97},
contributions coming from the attachment of photon to any charged internal 
line are strongly suppressed and we can neglect them safely. However,
since the mass of $\ell$-lepton is not much smaller than that of $B_s$-meson, 
in $B_s\to \ell^+\ell^-\gamma$ decay, the contributions of the 
diagrams with photon radiating from  final leptons are  comparable  with  
those from initial quarks.  When a photon is attached to the initial quark 
lines, the corresponding matrix element for the $B\to
\ell^+\ell^-\gamma$ decay can be written as 
\begin{eqnarray}
{\cal M}_1=\frac{\alpha_{em}^{3/2}G_F}{\sqrt{2\pi}}K_{tb}K_{ts}^*
&&\left\{[A\varepsilon_{\mu\alpha\beta\sigma}\epsilon_\alpha^*p_\beta q_\sigma
+iB(\epsilon_\mu^*(pq)-(\epsilon^*p)q_\mu)]\bar{\ell}\gamma_\mu\ell\right.
\nonumber\\
&&+\left.[C\varepsilon_{\mu\alpha\beta\sigma}\epsilon_\alpha^*p_\beta q_\sigma
+iD(\epsilon_\mu^*(pq)-(\epsilon^*p)q_\mu)]\bar{\ell}\gamma_\mu\gamma_5\ell
\right\},
\label{M1}
\end{eqnarray}
where 
\begin{eqnarray}
A&=&\frac{1}{m_{B_s}^2}\left[C_9^{eff}G_1(p^2)-2C_7^{eff}\frac{m_b}{p^2}G_2(p^2)
\right],\nonumber\\
B&=&\frac{1}{m_{B_s}^2}\left[C_9^{eff}F_1(p^2)-2C_7^{eff}\frac{m_b}{p^2}F_2(p^2)
\right]\nonumber\\
C&=&\frac{C_{10}}{m_{B_s}^2}G_1(p^2),\nonumber \\
D&=&\frac{C_{10}}{m_{B_s}^2}F_1(p^2).
\end{eqnarray}
In obtaining Eq.\ (\ref{M1}) we have used  
\begin{eqnarray}
\langle\gamma|\bar{s}\gamma_\mu(1\pm\gamma_5)|B_s\rangle
=&\frac{e}{m_{B_s}^2}\left\{\varepsilon_{\mu\alpha\beta\sigma}\epsilon_\alpha^*
p_\beta q_\sigma G_1(p^2)\mp i\left[(\epsilon_\mu^*(pq)-(\epsilon^*p)q_\mu)
\right]F_1(p^2)
\right\},\\
\label{gmb1} 
\langle\gamma|\bar{s}i\sigma_{\mu\nu}p_\nu(1\pm\gamma_5)b|B_s\rangle
=&\frac{e}{m_{B_s}^2}\left\{\varepsilon_{\mu\alpha\beta\sigma}\epsilon_\alpha^*
p_\beta q_\sigma G_2(p^2)\pm i\left[(\epsilon_\mu^*(pq)-(\epsilon^*p)q_\mu)
\right]F_2(p^2)
\right\},
\label{gmb2} 
\end{eqnarray}
and 
\begin{equation}
\langle\gamma|\bar{s}(1\pm\gamma_5)|B_s\rangle=0.
\label{gmb3} 
\end{equation}
Here $\epsilon_\mu$ and $q_\mu$ are the four vector polarization and momentum of 
photon, respectively; $G_i,\ F_i$ are form factors\cite{Buchalla93,Eilam95}. 
Eq.\ (\ref{gmb3}) can be obtained by multiplying\ $p_\mu$ in both sides of 
Eq.\ (\ref{gmb1}) and using the equations of motion.  From Eq.\ (\ref{gmb3}) 
one can see that the neutral scalars do not contribute to the matrix element 
${\cal M}_1$.

When a photon is radiated from the final $\ell$-leptons, the situation is 
different. Using the expressions
\begin{eqnarray}
& & \langle 0|\bar{s}b|B_s\rangle=0,\nonumber\\
& & \langle 0|\bar{s}\sigma_{\mu\nu}(1+\gamma_5)b|B_s\rangle =0,\nonumber\\
& & \langle 0|\bar{s}\gamma_\mu\gamma_5|B_s\rangle =-if_{B_s}P_{B_s\mu}
\end{eqnarray}
and the conservation of the vector current, one finds that only  
the operators ${\cal Q}_{1,2}$ and ${\cal O}_9$ give contribution to this 
Bremsstrahlung part. The corresponding matrix is given by \cite{Iltan00}
\begin{eqnarray}
{\cal M}_2&=&\frac{\alpha_{em}^{3/2}G_F}{\sqrt{2\pi}}K_{tb}K_{ts}^*
i2m_\ell f_{B_s}\left\{(C_{10}+\frac{m_{B_s}^2}{2m_\ell m_b}C_{Q_2})\bar{\ell}
\left[\frac{\not\epsilon\not P_{B_s}}{2p_1q}-
\frac{\not P_{B_s}\not\epsilon}{2p_2q}\right]\gamma_5\ell\right.\nonumber\\
&&+\left.\frac{m_{B_s}^2}{2m_\ell m_b}C_{Q_1}
\left[2m_\ell(\frac{1}{2p_1q}+\frac{1}{2p_2q})\bar{\ell}\not\epsilon\ell+
\bar{\ell}(\frac{\not\epsilon \not P_{B_s}}
{2p_1q}-\frac{\not P_{B_s}\not\epsilon}{2p_2q})\ell\right]\right\}.
\label{matrix2}
\end{eqnarray}
Here  $P_{B_s},\ f_{B_s}$ are the momentum and the decay constant of 
the $B_s$ meson, $p_1,\ p_2$ are momenta of the final $\ell$-leptons. 
 
Finally, the total matrix element for the $B_s\to \ell^+\ell^-\gamma$
decay is obtained as a sum of the ${\cal M}_1$ and ${\cal M}_2$. After 
summing over the spins of the $\ell$-leptons and polarization of the photon, 
we get the square of the matrix element as
\begin{equation}
\vert{\cal M}\vert^2=\vert{\cal M}_1\vert^2+\vert{\cal M}_2\vert^2+
2Re({\cal M}_1{\cal M}_2^*)
\end{equation}
with
\begin{eqnarray}
\vert{\cal M}_1\vert^2&=&
4\vert\frac{\alpha_{em}^{3/2}G_F}{\sqrt{2\pi}}K_{tb}K_{ts}^*\vert^2
\left\{\left[\vert A\vert^2+\vert B\vert^2\vert\right]
\left[p^2\left((p_1q)^2+(p_2q)^2\right)+2m_\ell^2(pq)^2\right]\right.\nonumber\\
&&+\left.\left[\vert C\vert^2+\vert D\vert^2\vert\right]
\left[p^2((p_1q)^2+(p_2q)^2)-2m_\ell^2(pq)^2\right]\right.\nonumber\\
&&\left.+2Re(B^*C+A^*D)p^2\left((p_1q)^2-(p_2q)^2\right)\right\},
\end{eqnarray}
\begin{eqnarray}
2Re({\cal M}_1{\cal M}_2^*)&=&
-16\vert\frac{\alpha_{em}^{3/2}G_F}{\sqrt{2\pi}}K_{tb}K_{ts}^*\vert^2
m_\ell^2f_{B_s}(pq)^2\left\{
\vert C_{10}+\frac{m_{B_s}^2C_{Q_2}}{2m_\ell m_b}\vert\left[Re(A)
\frac{(p_1q+p_2q)}{(p_1q)(p_2q)}\right.\right.\nonumber\\
&&\left.\left.-Re(D)\frac{(p_1q-p_2q)}
{(p_1q)(p_2q)}\right]+ Re(B)\vert\frac{m_{B_s}^2C_{Q_1}}{2m_\ell m_b}\vert
\left[\frac{3m_{B_s}^2+2m_\ell^2-5(pq)}{(p_1q)(p_2q)}-
\frac{2p^2}{(pq)^2}\right]\right.\nonumber\\
&&\left.+Re(C)\vert\frac{m_{B_s}^2C_{Q_1}}{2m_\ell m_b}\vert
\left[\frac{(p_1q-p_2q)}{(p_1q)(p_2q)}(1+\frac{2p^2}{(pq)^2})\right]\right\},
\end{eqnarray}
\begin{eqnarray}
\vert{\cal M}_2\vert^2&=&
-8\vert\frac{\alpha_{em}^{3/2}G_F}
{\sqrt{2\pi}}K_{tb}K_{ts}^*\vert^2m_\ell^2f_{B_s}^2
\left\{\vert C_{10}+\frac{m_{B_s}^2C_{Q_2}}{2m_\ell m_b}\vert^2
\left[\frac{m_\ell^2m_{B_s}^2(pq^2)}{(p_1q)^2(p_2q)^2}-
\frac{m_{B_s}^2p^2+2(pq)^2}{(p_q)(p_2q)}\right]
\right.\nonumber\\
&&\left.-\vert\frac{m_{B_s}^2C_{Q_1}}{2m_\ell m_b}\vert^2
\left[\frac{m_\ell^2(m_{B_s}^2-4m_\ell^2)(pq)^2}{(p_1q)^2(p_2q)^2}
-\frac{(m_{B_s}^2-4m_\ell^2)p^2+2(pq)^2}{(p_1q)(p_2q)}\right]
\right\}.
\end{eqnarray}
It is obvious that the quantity $\vert{\cal M}\vert^2$ depends only on the 
scalar products of the momenta of the external particles. 
In this paper, we follow Ref.\ \cite{Aliev97} and 
consider the photon in  $B_s\to\ell^+\ell^-\gamma$ 
as a hard photon and impose a cut on the photon energy $E_\gamma$
\footnote{When photon is soft, both processes of $B_s\to\ell^+\ell^-\gamma$ 
and $B_s\to\ell^+\ell^-$ should be considered together, and in this case, 
the infrared singular terms in $\vert{\cal M}_2\vert^2$ can be canceled exactly 
by the $O(\alpha_{em})$ virtual correction in $B_s\to\ell^+\ell^-$ 
\cite{Aliev97}.}, which 
correspond to the radiated photon can be detected in  the experiments. 
This cut requires  $E_\gamma\geq \delta~ m_{B_s}/2$ with $\delta = 0.02$.

After integrating over the phase space and the lepton 
energy $E_1$, we express the decay rate as
\begin{eqnarray}
\Gamma&=&\vert\frac{\alpha_{em}^{3/2}G_F}{2\sqrt{2\pi}}K_{tb}K_{ts}^*\vert^2
\frac{m_{B_s}^5}{(2\pi)^3}\left\{
\frac{m_{B_s}^2}{12}\int^{1-\delta}_{4\hat{r}}(1-\hat{s})^3d\hat{s}\sqrt{1-
\frac{4\hat{r}}{\hat{s}}} 
\left[\left(|A|^2+|B|^2\right)(\hat{s}+2\hat{r})\right.\right.\nonumber\\
&&\left.\left.+\left(|C|^2+|D|^2\right)(\hat{s}-4\hat{r})\right]
-2f_{B_s}\vert C_{10}+\frac{m_{B_s}^2C_{Q_2}}{2m_\ell m_b}\vert
\hat{r}\int^{1-\delta}_{4\hat{r}}(1-\hat{s})^2d\hat{s} Re(A)
\ln\ \hat{z}\right.\nonumber\\
&&-2f_{B_s}\vert\frac{m_{B_s}^2C_{Q_1}}{2m_\ell m_b}\vert
\hat{r}\int^{1-\delta}_{4\hat{r}}(1-\hat{s}) d\hat{s} Re(B)
\left[(1+4\hat{r}+5\hat{s})\ln\ \hat{z}+
\hat{s}\sqrt{1-\frac{4\hat{r}}{\hat{s}}}\right]\nonumber\\
&&\left.-\frac{4f_{B_s}^2}{m_{B_s}^2}
\vert C_{10}+\frac{m_{B_s}^2C_{Q_2}}{2m_\ell m_b}\vert^2\hat{r}
\int^{1-\delta}_{4\hat{r}}d\hat{s}\left[(1+\hat{s}+
\frac{4\hat{r}-2}{1-\hat{s}})\ln\ \hat{z}+
\frac{2\hat{s}}{1-\hat{s}}
\sqrt{1-\frac{4\hat{r}}{\hat{s}}}\right]\right.\nonumber\\
&&\left.+f_{B_s}^2
\vert\frac{C_{Q_1}}{m_b}\vert^2
\int^{1-\delta}_{4\hat{r}}d\hat{s}\left[(1-8\hat{r}+\hat{s}
-\frac{2-10\hat{r}+8\hat{r}^2}
{1-\hat{s}})\ln\ \hat{z}+\frac{2(1-4\hat{r})\hat{s}}{1-\hat{s}}
\sqrt{1-\frac{4\hat{r}}{\hat{s}}}\right]\right\} \ ,
\label{gbttg}
\end{eqnarray}
where $\hat{s}=p^2/m_{B_s}^2,\ \hat{r}=m_\ell^2/m_{B_s}^2$,  
$\hat{z}\equiv Z(\hat{r},\hat{s})$ takes the form given in Eq.\ (\ref{zxy}). 

\section{Experimental constraints on the relevant SUSY parameters}
\label{constraints}
Before scanning the relevant parameter space of MSSM, 
we assume 
(i) the masses of the particles are  
restricted to the sub-TeV regime and larger than the lower experimental
bounds\cite{PDG00}; 
(ii) A-parameters are smaller than $3M_{\tilde Q}$\cite{Haber92};
(iii) Except for the third generation squarks, all sfermions are degenerate
and have masses of $\sim 1 TeV$; 
(iv) The GUT mass relation $M_1\approx M_2/2$ for gauginoes is used; 
(v) Only small flavor violation $\epsilon_i<0.1$ is allowed.
We also take into account the well-known large radiative corrections to neutral 
Higgs masses \cite{Djouani97}.  With above assumptions and gluino mass fixed as 
its lower experimental bound $190~GeV$\cite{PDG00}, the relevant parameter space 
is determined by nine input parameters:  $M_{\tilde Q}$, $M_{{\tilde t}_R}$, 
$M_{{\tilde b}_R}$, $\mu$, $\tan\beta$, $A_t$, $A_b$, $M_2$ and $m_A$. 
In addition, we consider the following the experimental constraints in our scan: 

(1) The recently reported value of muon $g-2$ \cite{Brown01} shows a 2.6 standard 
deviation from its SM prediction. The SUSY explanation of this deviation 
requires (i) $\mu>0$ and  (ii) large $\tan\beta$. In our calculation
we assume $20\le\tan\beta\le 50$ and at least one of the charginos or neutralinos 
must be lighter than 500 GeV\cite{Everett00}.

(2) Non-observation of any supersymmetric signals at CERN $e^+e^-$ collider
LEP-II and the Fermilab Tevatron imposes lower bounds as
\begin{equation}
\begin{array}{ccc}
 m_{H^\pm}\geq 78.6~GeV & m_{h^0}\geq 88.3~GeV, 
    &m_{{\tilde\chi}^\pm_1}\geq 67.7~GeV,\\
 m_{{\tilde\chi}^0_1}\geq 42.0~GeV &m_{{\tilde t}_1}\geq 86.4~GeV,
 &m_{{\tilde b}_1}\geq 75.0~GeV.
\end{array}
\end{equation}

(3) The latest measurement of the inclusive branching ratio from CLEO
and BELLE \cite{CLEO01} gives world average value 
\begin{equation}
2.44\times 10^{-4}<{\em Br}(B\rightarrow {\em X}_s\gamma)
<4.02\times 10^{-4}  ~~ (95\% ~C.~L.) ,
\label{bsrex}
\end{equation}  
which is specially useful to constrain extensions of the SM. Previous studies
used leading-order (LO) SM result to limit the MSSM parameter space. Instead of 
LO calculation\cite{Grinstein88,Ciuchini93}, the branching ratio of
$B(B\rightarrow {\em X}_s\gamma)$ has been estimated at 
next-leading-order (NLO) level in the SM\cite{Chetyrkin97} 
with about 22.5\% increase of its central value and theoretical error less 
than 10\%. In this paper, we use the SM result computed at NLO level whereas 
the additional supersymmetric contributions at LO level.
We do not use the available NLO matching conditions for the supersymmetric particles
since they are computed under the specific assumptions about the sparticle 
spectrum, not necessarily satisfied in the criteria, and moreover, they are not 
valid for large values of $\tan\beta$\cite{Chankowski01}.  Since the Wilson coefficient 
$C_8$ accounts for only $3\%$ of the standard model $b\to s\gamma$ amplitude, it 
is therefore not expected to be significantly more important in the MSSM.  In 
this case, Eq. (\ref{bsrex}) implies the constraints on the ratio 
$R_7=C_{7,MSSM}^{0,eff}/C_{7,SM}^{0,eff}$
\begin{equation}
0.83\leq R_7\leq 1.13,  ~~or~~ -1.24\leq R_7\leq -0.94. 
\end{equation}

(4) The stringent bounds on the magnitude of the short distance 
coefficients come from the Collider Detector at Fermilab (CDF)\cite{PDG00} 
\begin{eqnarray}
&&Br(B^+\to K^+\ell^+\ell^-)<(60, 5.2)\times 10^{-6} ~~~(90\% C.L.),\nonumber\\
&&Br(B^0\to K^{*0}\ell^+\ell^-)<(290, 4.0)\times 10^{-6}~~ (90\% C.L.),\nonumber\\
&&Br(B_s\to ~~\ell^+\ell^-~~)~<~(5.4, 2.0)\times 10^{-6}~~ (95\% C.L.)
\end{eqnarray}   
for $\ell=e,\mu$. 

\section{Numerical results and discussions}
\label{sec:result}

We perform a complete scan over the  nine-dimensional 
parameter space of the MSSM. For reference, 
we present our SM predictions
\begin{eqnarray}
Br(B\to X_s\ell^+\ell^-)     &=&(11.8, 9.87,2.94)\times 10^{-6},\\
Br(B_s\to\ell^+\ell^-\gamma)&=&(2.50,2.62,5.49)\times 10^{-8} 
\end{eqnarray}
for $\ell=e,\mu, \tau$. These values are obtained by taking the QCD coupling constant 
$\alpha_s(m_b)=0.218$, the masses, decay widths and branching ratios of 
$J/\Psi$ family in Ref.\ \cite{PDG00}, the normalized factor $Br(B\to X_c\ell\nu)=10.2\%$ 
and the form-factors\cite{Eilam95}
\begin{eqnarray}
G_1(p^2)&=&\frac{1\ GeV}{(1-p^2/5.6^2)^2},\  \ \ \  
G_2(p^2)=\frac{3.74\ GeV}{(1-p^2/40.5)^2},\nonumber\\
F_1(p^2)&=&\frac{0.8\ GeV}{(1-p^2/6.5^2)^2}, \ \ \ \
F_2(p^2)=\frac{0.68\ GeV}{(1-p^2/30)^2} . 
\label{formf}
\end{eqnarray}
as well as  the fixed input parameters \cite{PDG00} listed in Table I 
in the numerical calculations.

\begin{figure}[htb]
\begin{center}
\epsfig{file=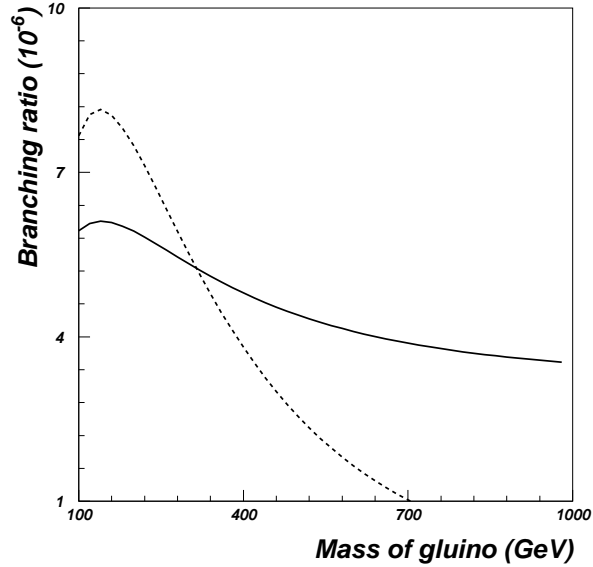,width=8cm}              
\caption{The branching ratios vs gluino mass
with $\tan\beta=40$, $m_A=271~GeV$, $m_{\tilde{t}_1}=320~GeV, 
m_{\tilde{b}_1}=265~GeV$. The solid and dashed lines 
correspond to the inclusive and exclusive decays with $\ell=\tau$, respectively.}
\label{brg}
\end{center}
\end{figure}

\null\vspace{0.4cm}
\noindent
{\small Table I. 
  The value of the input parameters used in the numerical calculations
 (mass and decay constant in unit GeV)}.
\begin{center}
\begin{tabular}{cccccc}  
\hline
$m_t$& $m_c$&$m_b$&$m_\tau$ & $m_{B_s}$& $m_W$\\
\hline
176  & 1.4 & 4.8 & 1.78 & 5.26  & 80.45 \\
\hline \hline								
$f_{B_s}$&~~$|K_{tb}K_{ts}^*|$~~ & 
~~$|K_{tb}K_{ts}^*/K_{cb}|^2$~~
&$\alpha_{em}^{-1}$&$\tau(B_s)$&$\sin^2\theta_W$\\
\hline
0.21& 0.045&0.95&137&$~~1.64\times 10^{-12} s~~$&0.233\\ \hline
\end{tabular}
\end {center}
\begin{figure}[htb]
\begin{center}
\epsfig{file=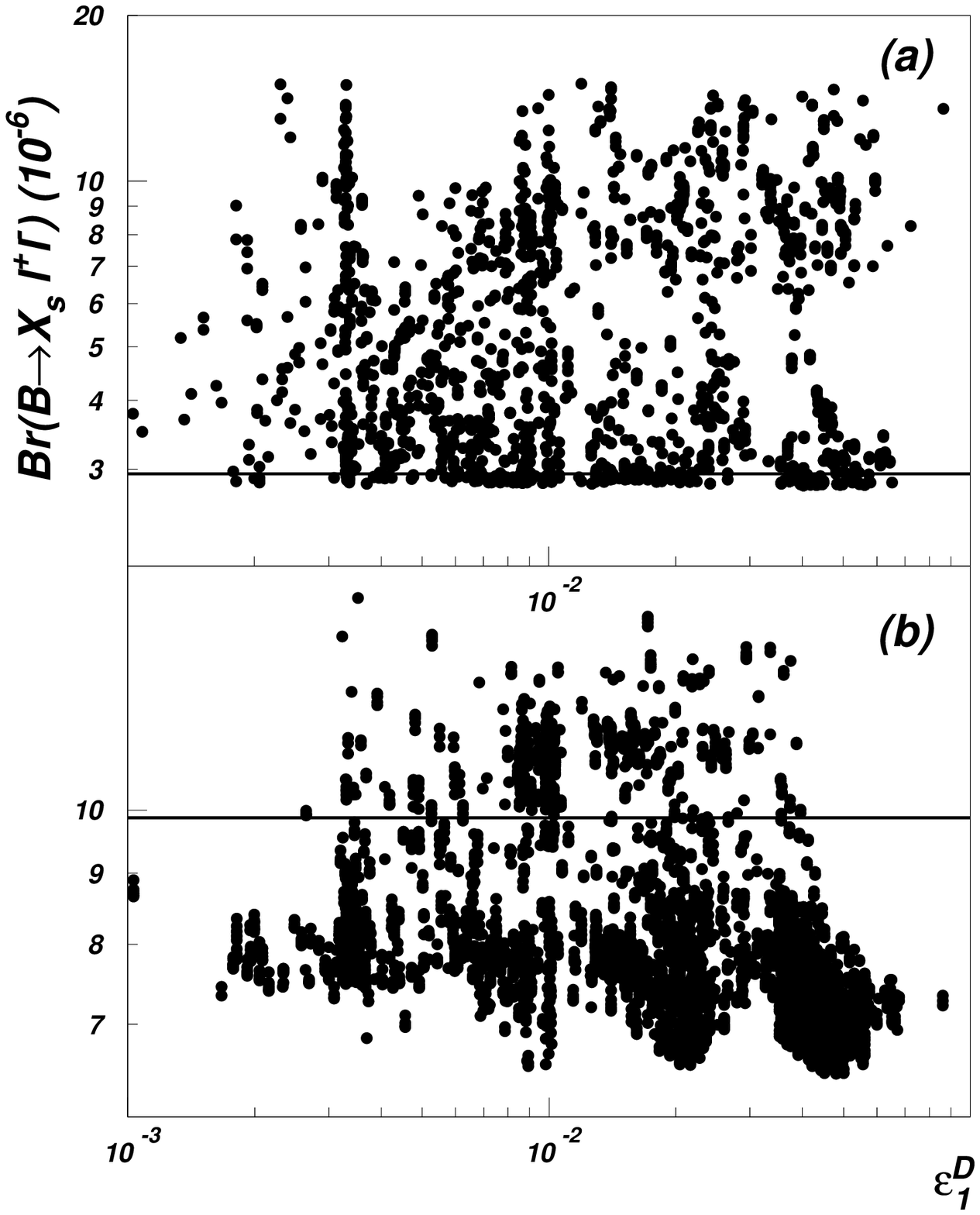,width=8cm}
\epsfig{file=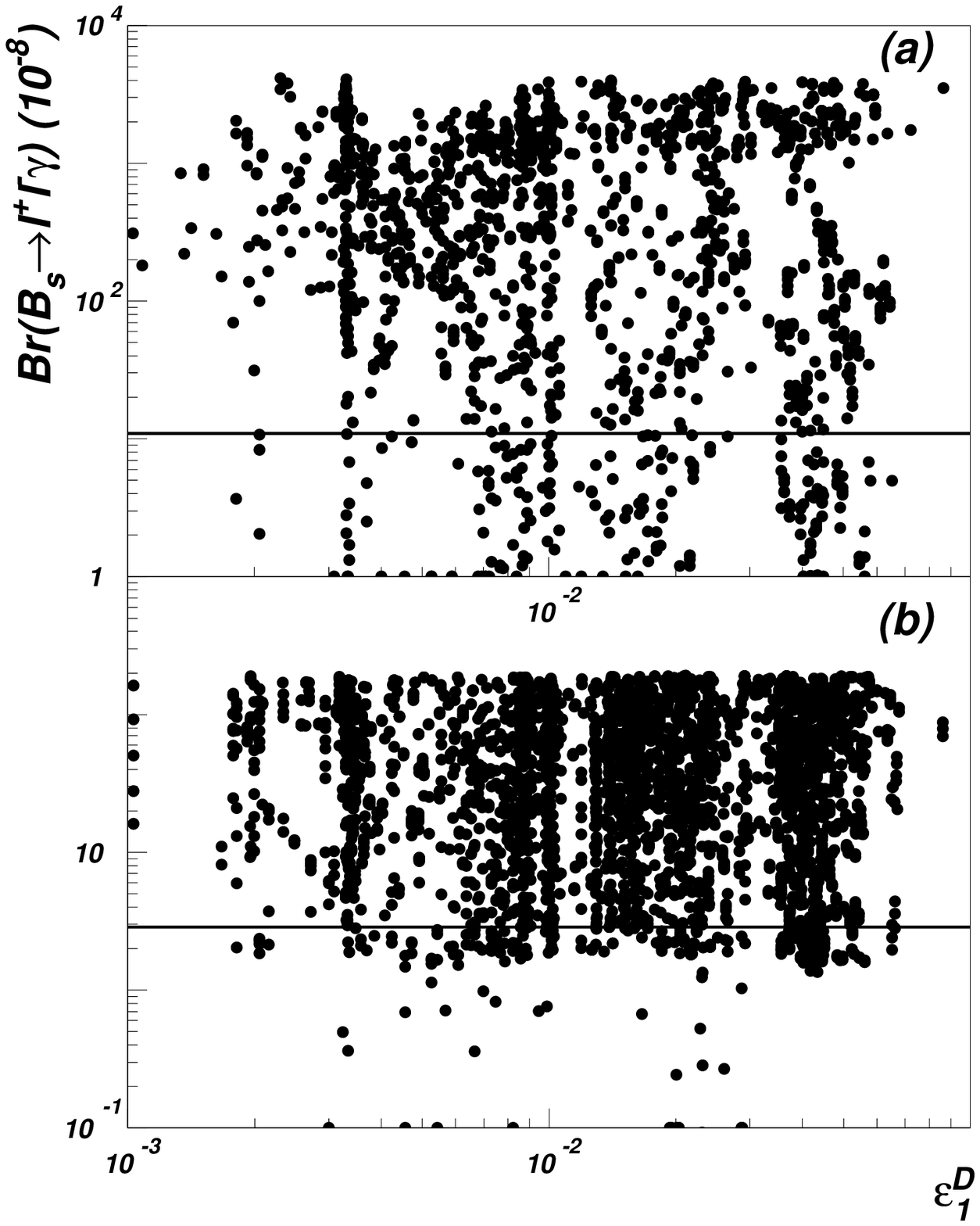,width=8cm}
\caption{The parameter space scatter plot of $Br(B\to X_s\ell^+\ell^-)$
(left) and $Br(B_s\to\ell^+\ell^-\gamma)$ (right) vs $\epsilon_1$ for 
$\ell=\tau$ in (a) and for $\ell=\mu$ in (b). 
The solid lines stand for the SM predictions }
\label{bttep1}
\end{center}
\end{figure}
The results shown in Fig.~\ref{bttep1} indicates that in some part of parameter space,  
the enhancement factors $R$ of  the branching ratios can be 5
 for the inclusive decay $B\to X_s\tau^+\tau^-$ 
and a couple of orders of magnitude over the standard model predictions 
for the exclusive decay $B_s\to\tau^+\tau^-\gamma$. 
This is quite different from the previous studies\cite{Huang99}. 
As illustrative examples, the branching ratios for ditau final state as functions
of the gluino mass are plotted in Fig.~\ref{brg}. The effects drop with the 
increase of gluino mass, showing the decoupling property of the MSSM.
Table 2 shows the case in which both enhancement factors reach their 
maximum values and the corresponding SUSY parameters.
A comparison of the enhancement factors with and without the NCL effects 
is also presented.  

\null
\vspace{0.2cm}
\noindent
{\small Table II.
The maximum enhancement factors and their corresponding SUSY parameters
(mass in unit GeV).}
\begin{center}
\begin{tabular}{c|c|c|c}
\hline
\multicolumn{2}{c|}{$R_{max}(B\to X_s\tau\tau)$}
&\multicolumn{2}{c}{$R_{max}(B_s\to\tau\tau\gamma)$}\\
\hline
With NCL & without NCL&With NCL & without NCL\\
4.34& 0.99& 327.0&0.75\\
\hline
 \multicolumn{4}{c}{$\tan\beta=40$~~  $m_A=453$~~   
 $m_{\tilde{t}_1}=143$~~ $m_{\tilde{b}_1}=92$}\\
\hline\hline
\end{tabular}
\end {center}
\vspace{0.2cm}
We have the following comments on the results:

(1) The constraint on $B_s\to\ell^+\ell^-$ is of special useful to 
limit the effects of the scalars and pseudo-scalars. However,
since no helicity suppression compared with $C_{10}$ (see Eq.~\ref{matrix2}), 
the contribution of $C_{Q_{1,2}}$ in $ B\to X_s\tau^+\tau^-$ and 
$B_s\to\ell^+\ell^-\gamma$ decay is still dominate for large $\tan\beta$. 

(2) As helicity suppression, the effects of $C_{Q_{1,2}}$ in     
$B\to X_s\tau^+\tau^-$ are very different from those with other dilepton 
final states (see Eq.~\ref{bllp}). The smaller mass of lepton in the final 
states, the less effect of gluino and the neutralinos. In the inclusive 
decays with lighter dilepton final states, the dominate contribution
comes from the interaction term $2Re(C_7^{eff}C_9^{eff})$ with opposite
sign to its SM value.  For large $\tan\beta$ region, 
relative large $m_\tau$ make the neutral Higgs with NCL contribute 
sizably to the inclusive decay.

\begin{figure}[htb]
\begin{center}
\epsfig{file=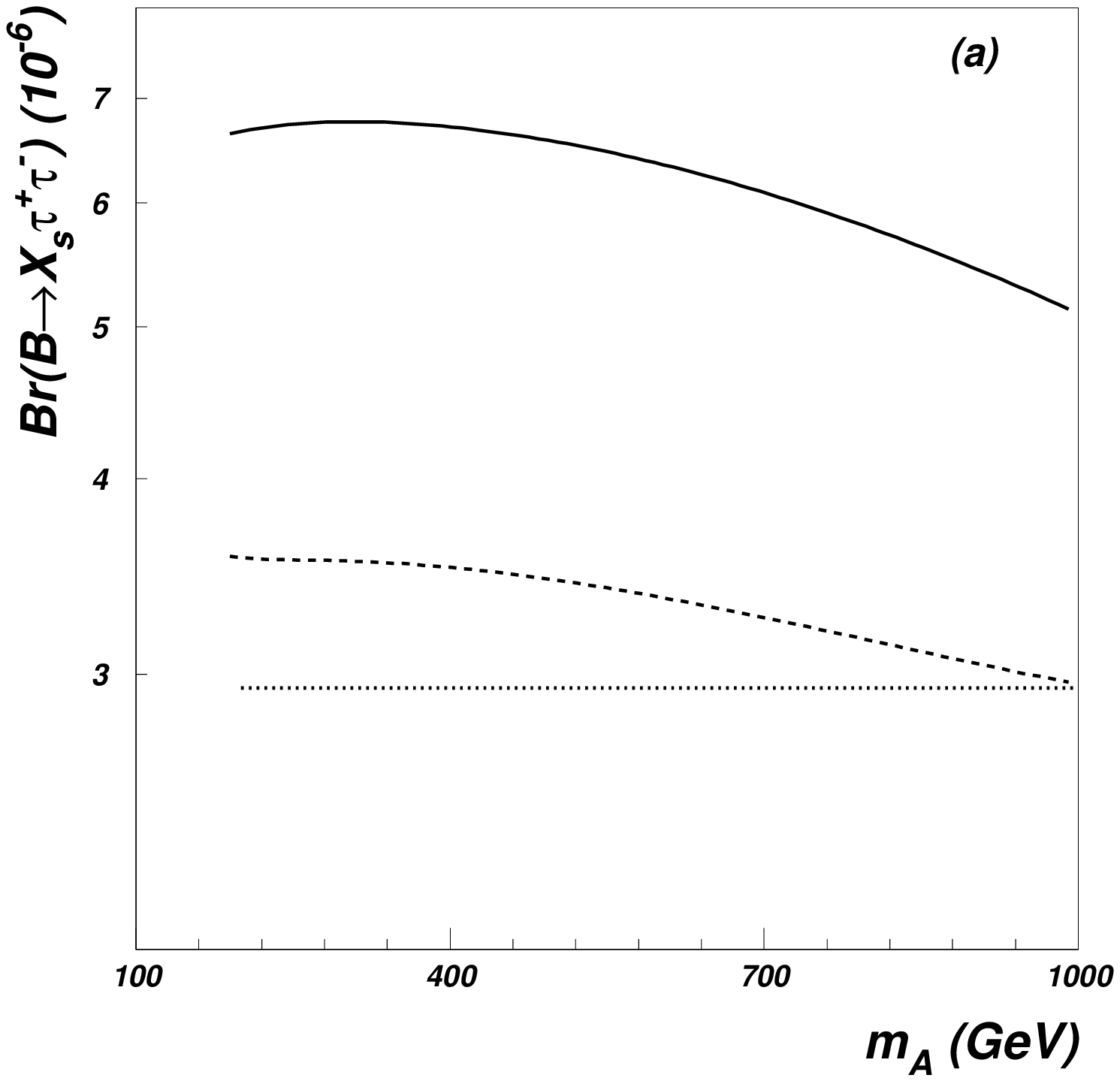,width=8cm}
\epsfig{file=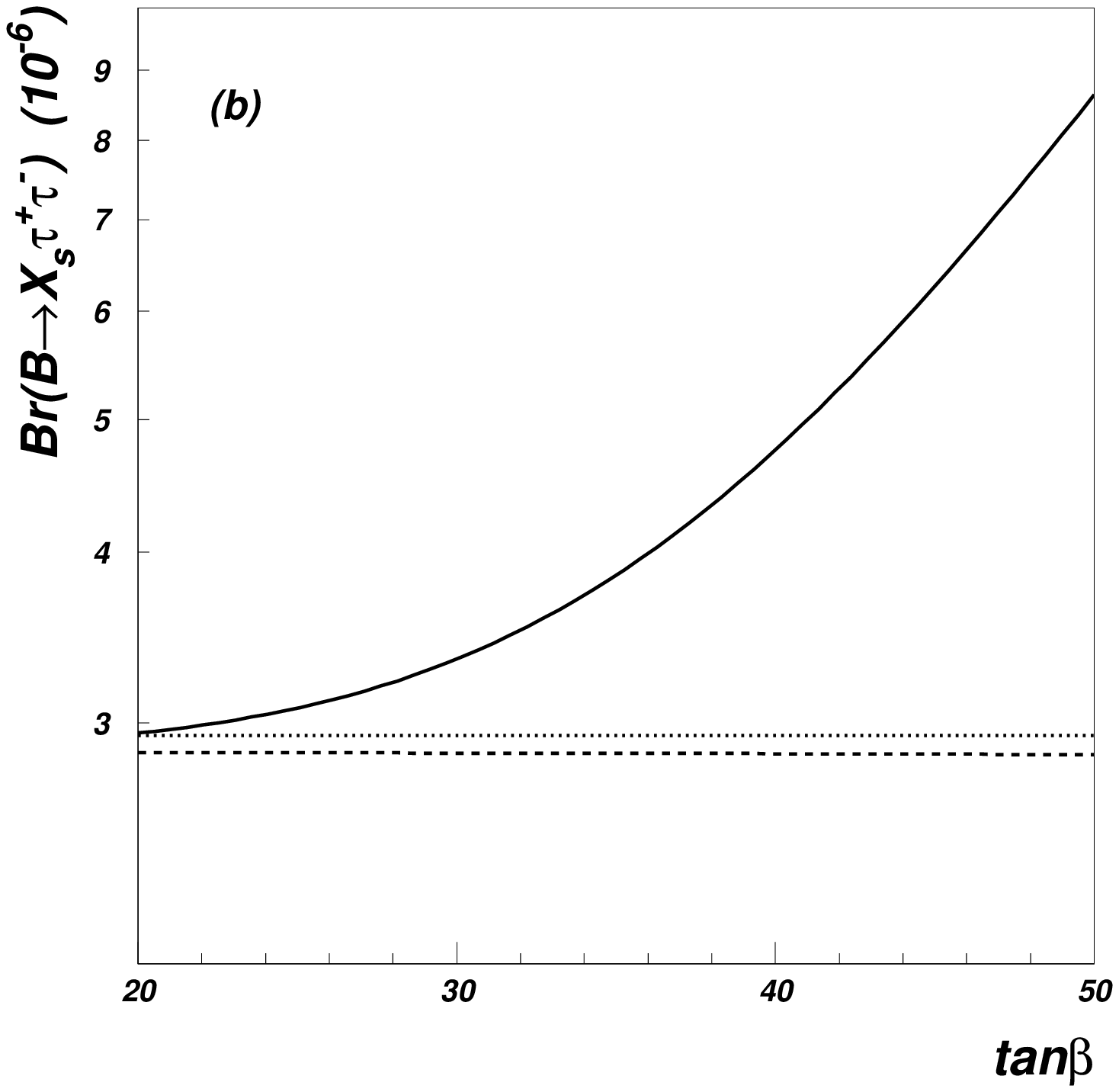,width=8cm}              
\caption{Branching ratio of $B\to X_s\tau^+\tau^-$ versus $m_A$ in (a) and
versus $\tan\beta$ in (b). The dotted line stands for the SM prediction, whereas 
the solid (dashed) one is the MSSM prediction with (without) the contributions of 
the gluino and neutralinos.  The parameter space in (a) is specified as  
$m_{\tilde t}=(179,304)~GeV, m_{\tilde b}=(257,700)~GeV$,
$m_{{\tilde\chi}^\pm}=(94,417)~GeV, m_{{\tilde\chi}^0}=(81,109,200,417)~GeV$ 
and $\tan\beta=20$; In (b), $m_A=899~GeV$, the masses of the sparticles
are dependent on the value of $\tan\beta$.}     
\label{btt}
\end{center}
\end{figure}

\begin{figure}
\begin{center}
\epsfig{file=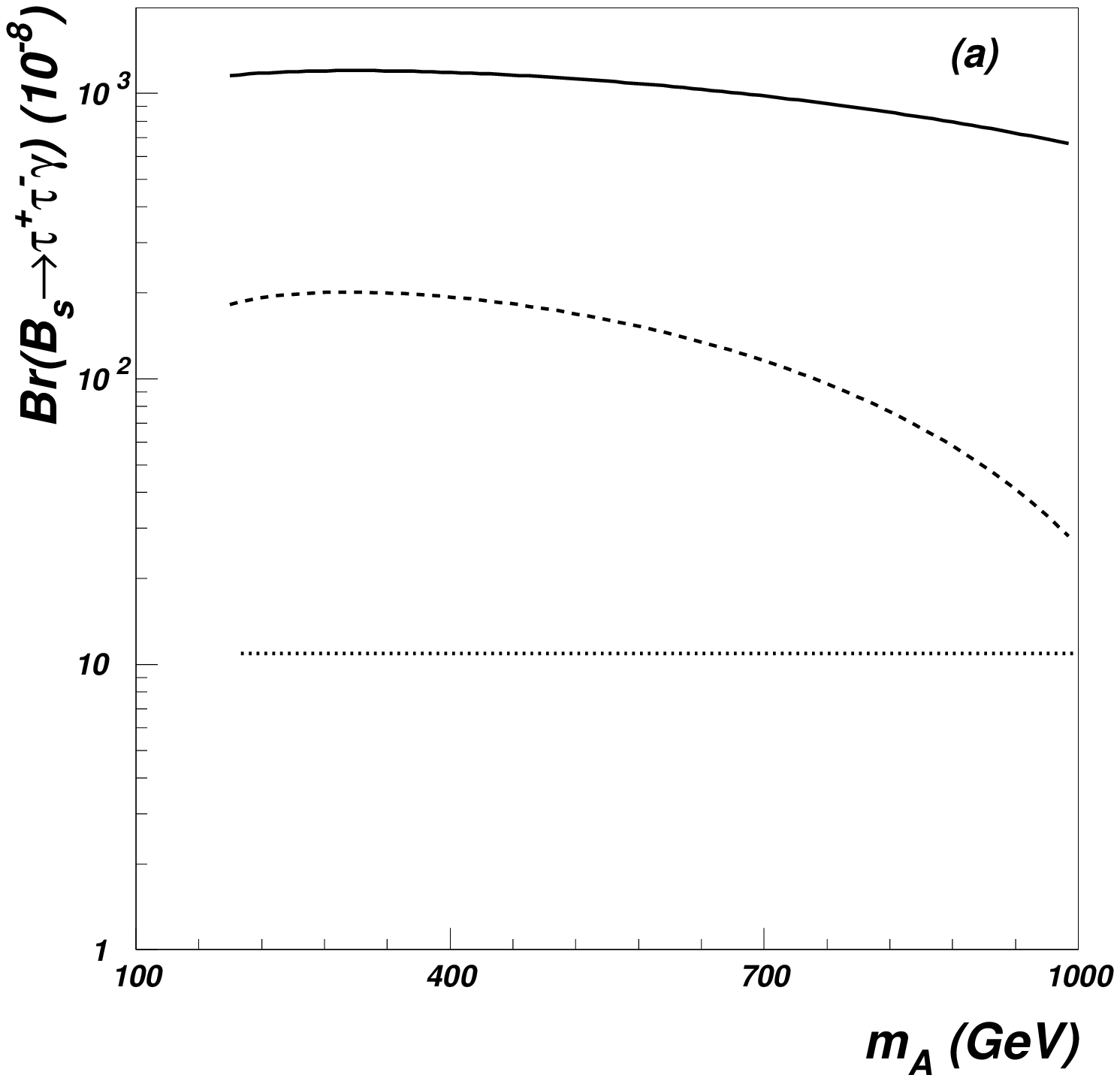 ,width=8cm}
\epsfig{file=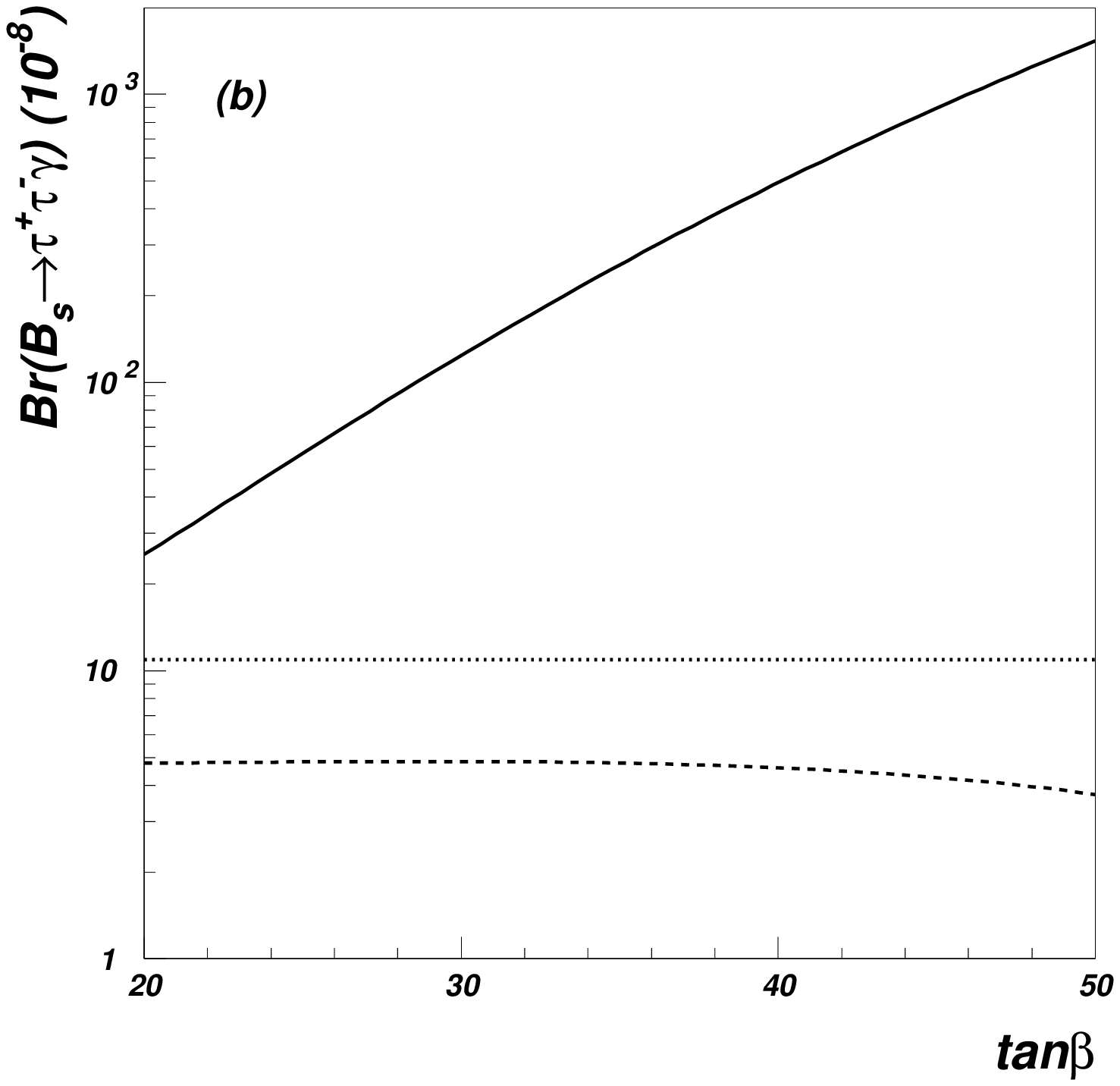 ,width=8cm}              
\caption{The same as Fig.~\ref{btt}, but for $B_s\to\tau^+\tau^-\gamma$.}
\label{bttg}
\end{center}
\end{figure}

\begin{figure}
\begin{center}
\epsfig{file=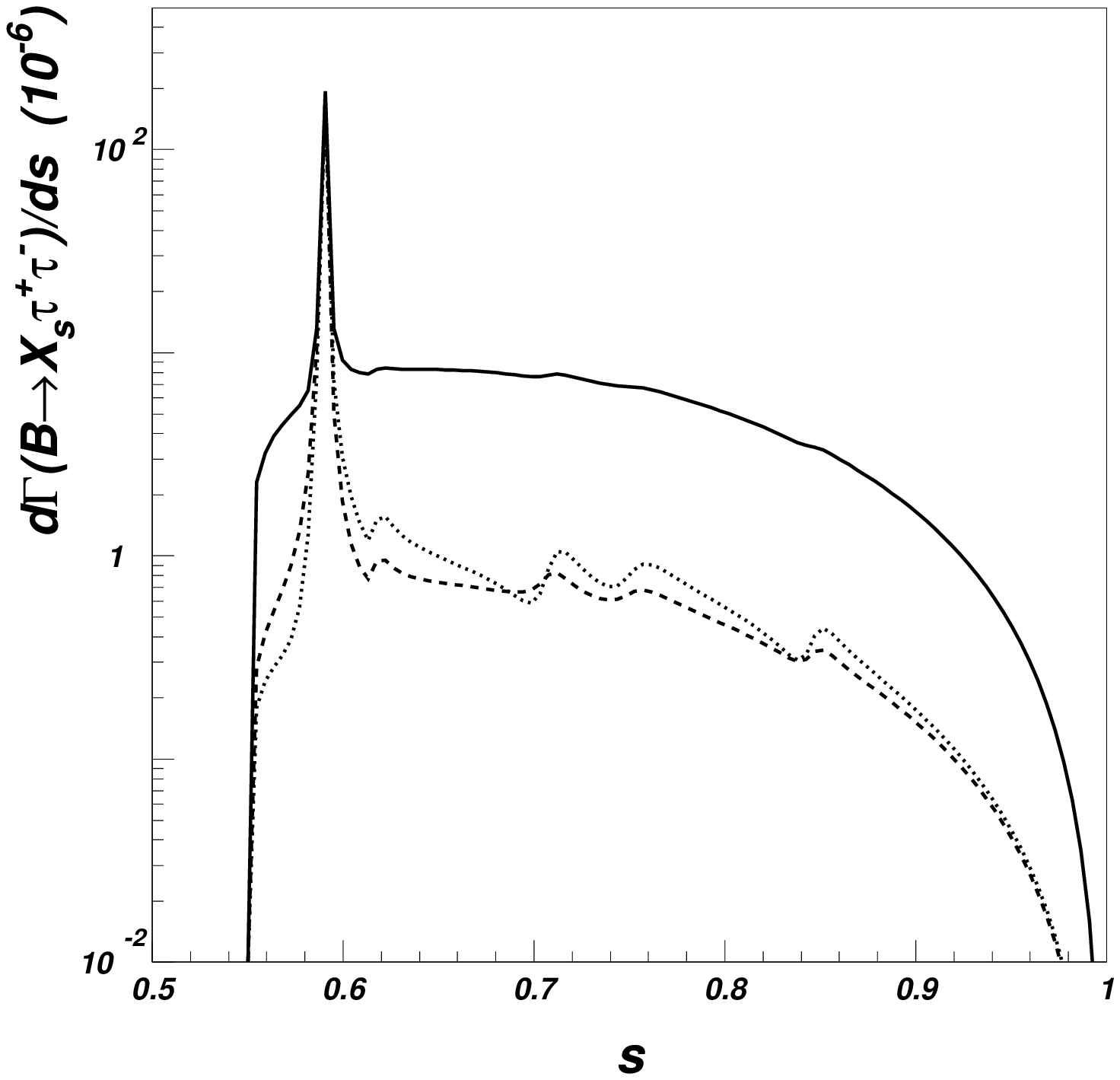 ,width=8cm} 
\epsfig{file=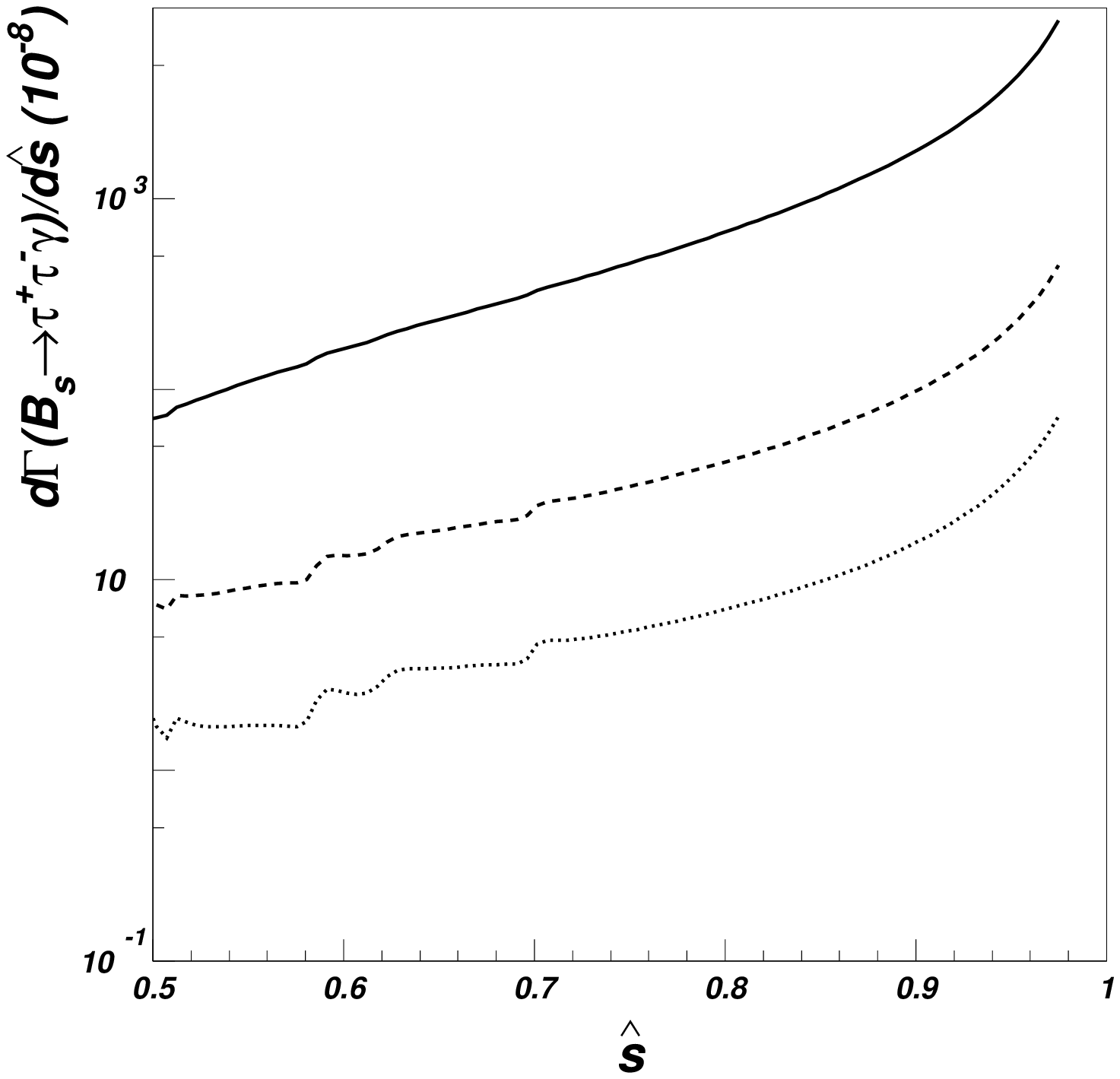 ,width=8cm} 
\caption{ The same as Fig.~\ref{btt}a, but for the differential branching 
ratios of $B\to X_s\tau^+\tau^-$ and $B_s\to\tau^+\tau^-\gamma$
versus the scaled invariant dilepton mass squared  with $m_A=305~GeV$.}
\label{dbtt-s}
\end{center}
\end{figure}

Some numerical examples in the region of $R_7>0$ are presented
in Figs. \ref{btt}-\ref{dbtt-s}.
Fig.~\ref{btt} and Fig.~\ref{bttg} show the branching ratios of
$B\to X_s\tau^+\tau^-$ and $B_s\to\tau^+\tau^-\gamma$,
respectively. For the specified parameter values in the figures,
one sees that the contributions from gluino and neutralinos are 
dominant. When the mass of the CP-odd Higgs is less than 700~GeV or 
$\tan\beta>40$, the branching ratio of $B\to X_s\tau^+\tau^-$ 
can be enhanced by a factor 2, and for  $B_s\to\tau^+\tau^-\gamma$,
by one order over the SM results.

\begin{figure}
\begin{center}
\epsfig{file=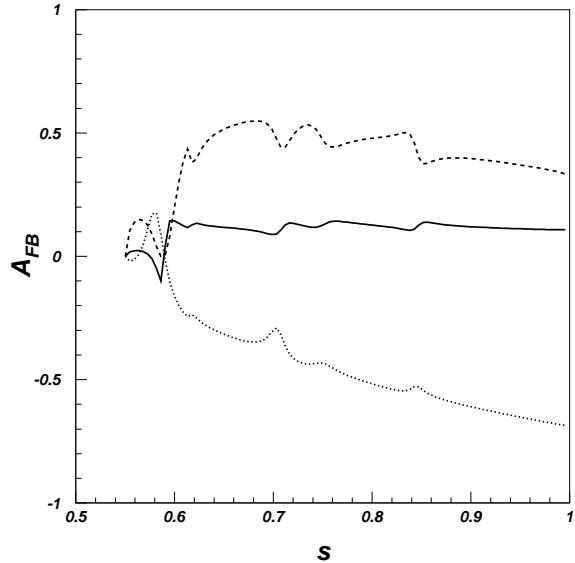,width=8cm}  
\caption{The same as Fig.~\ref{btt}a, but for the forward-backward asymmetry
of $B\to X_s\tau^+\tau^-$ versus the scaled invariant dilepton mass squared
$s$ with $m_A= 305~GeV$.}
\label{af-s}
\end{center}
\end{figure}

The differential branching ratios of  $B\to X_s\tau^+\tau^-$ and
$B_s\to\tau^+\tau^-\gamma$ versus the scaled invariant dilepton mass
squared are plotted in Fig.~\ref{dbtt-s}, while the dependence of the FB 
asymmetry for $B\to X_s\tau^+\tau^-$ on the scaled invariant dilepton mass
squared is shown in Fig.~\ref{af-s}. The figures show significant differences between 
the SM and the MSSM predictions, especially in large invariant dilepton mass region.  

We should point out that since  some common contributions appearing in both 
the numerator and the denominator cancel out to some extent, the FB asymmetry is a sensitive, 
relatively model-independent probe of these models. 
We stress that all these distributions would be useful for fitting the 
future experimental results in the framework of the MSSM, especially when 
some deviations from the SM predictions are discovered in future experiments. 
Different models, such as the MSSM and 2HDM's, may all predict some enhancements, 
but they may give different behaviors for some distributions. 
Although decay modes with the ditau final states 
are experimentally difficult compared to their di-muonic counterparts,
the considered decay modes are more sensitive to new physics. More 
theoretical studies, such as higher order effects which 
might also lead to important modifications as in the case of  
$B\to X_s\gamma$\cite{Degrassi00}, and experimental efforts for the 
decays considered are valuable.

\section{Conclusion}
\label{conclusion}

In this work we performed a complete one-loop calculation 
of the inclusive decay $B\to X_s\ell^+\ell$ and exclusive decay 
$B_s\to\ell^+\ell\gamma$ in the MSSM. Various  experimental 
constraints on the relevant SUSY parameters, such as 
$B\to X_s\gamma$, $B\to K^{(*)}\ell^+\ell^-, B_s\to\ell^+\ell^-$
and the latest $g_\mu-2$ experimental measurement, 
were considered to constrain the parameter space of the  MSSM. 
Our results showed that the contributions from the gluino and neutralino 
loops, which were neglected in previous studies, might be quite important 
or even  dominant in some part of parameter space. 
These supersymmetric contributions could  significantly enhance the branching 
ratios over the  SM predictions.
Also, with these contributions the  distributions of the forward-backward asymmetry 
of $B\to X_s\tau^+\tau^-$  and some other distributions could differ significantly 
from their SM predictions.  Up to now the currently running B-factories such as $BaBar$
at SLAC\cite{babar01} and BELLE at KEKB\cite{Belle01} have 
collected about $3.2\times 10^7~B\overline{B}$ pair with a luminosity of 
$(3\sim 4)\times 10^{33}~cm^{-2}s^{-1}$\cite{babar01,Belle01}. And 
the  $BaBar$ will take $10^8~B\overline{B}$ pairs in three years 
while the  BELLE $10^7~B\overline{B}$ pairs each year. 
Since the branching ratios of the inclusive decay $ B\to X_s\tau^+\tau^-$ 
and exclusive decay $B_s\to\tau^+\tau^-\gamma$ can be enhanced by the SUSY 
effects with large $tan\beta$ to reach the level of $10^{-5}$, these 
ditau decays might be observable at the B factories.
 
\section*{Acknowledgment}
We would like thank Prof. A.~Ali and C.-P. Yuan for useful discussion.   
This work is supported in part by a grant of Chinese Academy of Science
for Outstanding Young Scholars.

\section*{Appendix}

In the Appendix we present the operator basis $O_i,~Q_i$ and one-loop integral 
functions for $b\to s\ell^+\ell^-$.  

The operator basis $O_i$ is the same as the one used for the $b\to s\ell^+\ell^-$ in the SM,
\begin{eqnarray}
{\cal O}_1&=&(\bar{s}_\alpha\gamma^{\mu}Lc_\beta)
             (\bar{c}_\beta\gamma^{\mu}Lb_\alpha),\nonumber\\
{\cal O}_2&=&(\bar{s}_\alpha\gamma^{\mu}Lc_\alpha)
             (\bar{c}_\beta\gamma^{\mu}Lb_\beta),\nonumber\\
{\cal O}_{3,5}&=&(\bar{s}_\alpha\gamma^{\mu}Lc_\alpha)
            \left(\sum\limits_{q}\bar{q}_\beta\gamma^{\mu}(L,R)b_\beta\right),
              \nonumber\\
{\cal O}_{4,6}&=&(\bar{s}_\alpha\gamma^{\mu}Lc_\beta)
            \left(\sum\limits_{q}\bar{q}_\beta\gamma^{\mu}(L,R)b_\alpha\right),
              \nonumber\\
{\cal O}_7&=&\frac{e}{16\pi^2}
\bar{s}_\alpha\sigma^{\mu\nu}(m_bR+m_sL)b_\alpha F_{\mu\nu},\nonumber\\
{\cal O}_8&=&\frac{g_s}{16\pi^2}\bar{s}_\alpha\sigma^{\mu\nu}
(m_bR+m_sL)t_{\alpha\beta}b_\beta G_{\mu\nu}^\alpha,\nonumber\\
{\cal O}_9&=&\frac{e}{16\pi^2}(\bar{s}_\alpha\gamma^{\mu}Lb_\alpha)
(\bar{\ell}\gamma_\mu\ell),\nonumber\\
{\cal O}_{10}&=&\frac{e}{16\pi^2}(\bar{s}_\alpha\gamma^{\mu}Lb_\alpha)
(\bar{\ell}\gamma_\mu\gamma_5\ell),
\label{operator1}
\end{eqnarray}
where the chiral structure is specified by the projectors 
$L,R=(1\mp\gamma_5)/2$, while $\alpha$ and $\beta$ are color indices. 
$F_{\mu\nu}$ and $ G_{\mu\nu}^\alpha$
denote the QED and QCD field strength tensors, respectively.
$t_{\alpha\beta}$ are the color triplet generators,  $g$ and 
$g_s$ stand for the electromagnetic and strong coupling  constants.
 
Operators $Q_i$ come from exchanging the neutral Higgs bosons in MSSM
and are defined 
by \cite{Dai97}
\begin{eqnarray}
{\cal Q}_1&=&\frac{g^2}{16\pi^2}(\bar{s}_\alpha Rb_\alpha)
(\bar{\ell}\ell),\nonumber\\
{\cal Q}_2&=&\frac{g^2}{16\pi^2}(\bar{s}_\alpha Rb_\alpha)
(\bar{\ell}\gamma_5\ell),\nonumber\\
{\cal Q}_{3,4}&=&\frac{g^2}{16\pi^2}(\bar{s}_\alpha Rb_\alpha)
\left(\sum\limits_q\bar{s}_\beta(R,L)b_\beta\right),\nonumber\\
{\cal Q}_{5,6}&=&\frac{g^2}{16\pi^2}(\bar{s}_\alpha Rb_\beta)
\left(\sum\limits_q\bar{s}_\beta(R,L)b_\alpha\right),\nonumber\\
{\cal Q}_{7,8}&=&\frac{g^2}{16\pi^2}\bar{s}_\alpha\sigma^{\mu\nu}Rb_\alpha
\left(\sum\limits_q\bar{s}_\beta(R,L)b_\beta\right),\nonumber\\
{\cal Q}_{9,10}&=&\frac{g^2}{16\pi^2}\bar{s}_\alpha\sigma^{\mu\nu}Rb_\beta
\left(\sum\limits_q\bar{s}_\beta(R,L)b_\alpha\right).
\label{operator2}
\end{eqnarray}

The one-loop integral functions which appear within the MSSM matching 
conditions are given by 
\begin{eqnarray}
P_1(x)&=&\frac{1}{x-1}\ln x,\nonumber\\
P_2(x,y)&=&\frac{1}{(x-y)}
\left[\frac{x^2}{x-1}\ln x-\frac{y^2}{y-1}\ln y\right],\nonumber\\
P_3(x,y)&=&\frac{1}{(x-y)}\left[\frac{x}{x-1}\ln x-\frac{y}{y-1}\ln y\right],\nonumber\\
P_4(x,y,z)&=&\left[\frac{x}{(x-1)(x-y)(x-z)}\ln x+(x\leftrightarrow y)+
(x\leftrightarrow z)\right].
\end{eqnarray}

\end{document}